\documentclass[onecolumn,prb,aps,showpacs,superscriptaddress,floatfix]{revtex4-1}
\pdfoutput=1
\usepackage{graphicx}
\usepackage[colorlinks=true , citecolor=blue,urlcolor=blue]{hyperref}
\usepackage{color}
\usepackage{lmodern} 

\usepackage{amsmath}

\usepackage[english]{babel}
\usepackage{makecell}

\definecolor{bluegray}{rgb}{0.4, 0.6, 0.8}
\newcommand{\bluetitle}{\color{bluegray}}

\newcommand{\affA}{CNRS, Aix-Marseille Universit\'{e}, IM2NP (UMR 7334), Institut Mat\'{e}riaux Micro\'{e}lectronique et Nanosciences de Provence,  Marseille, France.}
\newcommand{\affB}{Department of Physics, The National High Magnetic Field Laboratory, Florida State University, Tallahassee, Florida 32310, USA.}
\newcommand{\affC}{CNRS, Universit\'{e} de Lille, LASIR (UMR 8516), Laboratoire de Spectrochimie Infrarouge et Raman, Villeneuve d'Ascq, France}
\newcommand{\affD}{Zernike Institute for Advanced Materials, University of Groningen, Nijenborgh 4, NL-9747 AG Groningen, The Netherlands}

\newcommand{\SI}{
	\setcounter{table}{0}
	\renewcommand{\thetable}{S\arabic{table}}
	\setcounter{figure}{0}
	\renewcommand{\thefigure}{S\arabic{figure}}
	\setcounter{equation}{0}
	\renewcommand{\theequation}{S\arabic{equation}}
}

\newcommand{\TB}[1]{\textbf{#1}}
\newcommand{\TI}[1]{\textit{#1}}

\newcommand{\mean}[1]{\left\langle #1\right\rangle}

\newcommand{\Gd}{CaWO$_4$:Gd$^{3+}$}
\newcommand{\Mn}{MgO:Mn$^{2+}$}
\newcommand{\DP}{IF(\Delta,\phi)}

\newcommand{\1}{\textcolor{red}} 

\begin{document}

	\title{Experimental protection of quantum coherence by using a phase-tunable image drive }

\author{S. Bertaina}\email{sylvain.bertaina@im2np.fr}\affiliation{\affA}
\author{H. Vezin}\affiliation{\affC}
\author{H De Raedt}\affiliation{\affD}
\author{I. Chiorescu} \email{ic@magnet.fsu.edu}\affiliation{\affB}


\begin{abstract}  
			The protection of quantum coherence is essential for building a practical quantum computer able to manipulate, store and read quantum information with a high degree of fidelity. Recently, it has been proposed to increase the operation time of a qubit by means of strong pulses to achieve a dynamical decoupling of the qubit from its environment. We propose and demonstrate a simple and highly efficient alternative route based on Floquet modes, which increases the Rabi decay time ($T_R$) in a number of materials with different spin Hamiltonians and environments. We demonstrate the regime $T_R \approx T_1$, thus providing a route for spin qubits and spin ensembles to be used in quantum information processing and storage.	
\end{abstract}

\maketitle
	\section{Introduction}
In open quantum systems, coherence of spin qubits is limited by spin-spin
interactions, spin diffusion, inhomogeneity of the static and microwave fields
\cite{Chirolli2008} as well as charge noise \cite{Yoneda2018a}. An increase in
coherence time is achieved by dynamically decoupling (DD) qubits from their
surroundings using distinct Electron Spin Resonance (ESR) pulses
\cite{Viola1998,Viola1999}. However, such pulses have inherent imperfections and
fluctuations, thus requiring their own layer of DD, resulting in a doubly
dressed qubit. The technique of concatenated DD  \cite{Cai2012,Cohen2017} has
been proposed for NV centers up to the second order of dressing
\cite{Cai2012,Farfurnik2017,Teissier2017,Rohr2014}. Here we demonstrate a pulse
protocol based on Floquet modes which successfully increases the decoherence
time in a number of materials with different spin Hamiltonians and environments,
such as low and high spin-orbit coupling for instance. Rather than focusing on
decoupling from the bath by strong excitation, we use very weak pulses and alter
the dynamics of the entire system. For short spin relaxation times accessible to our
measurement setup (at around 40 K) one can do a direct comparison with the
coherence time, and we demonstrate the regime $T_R\approx T_1$. In magnetic diluted systems $T_1\gg T_2$, \emph{e.g.}  $T_1$ of several $ms$ in rare earth ions such as Y$_2$SiO$_5$:Er$^{3+}$ \cite{Welinski2019} and Y$_2$SiO$_5$:Yb$^{3+}$ \cite{Lim2018} or $^{28}$Si:Bi with a tunable $T_1$ of thousand of seconds\cite{Bienfait2016}. Our general method can thus
lead to very long persistent Rabi oscillations, using a single circularly
polarized image pulse.

The use of strong continuous microwave excitation has been proposed as a way to
protect qubits \cite{Facchi2004,Fanchini2007} although the quantum gates would
need proper redesigning. In related studies, complex pulse design using an
arbitrary waveform generator, proved essential in studying Floquet Raman
transitions\cite{Shu2018,Saiko2018} and quantum metric of a two-level system
\cite{Yu2020} in NV centers. It is worth noting that in the case of concatenated
DD, the frequency of the second order ($n=2$) excitation has to match the Rabi
frequency of the first excitation ($n=1$); also, the two excitations are
linearly polarized and perpendicular to each other (the method extends to higher
orders in $n$). Experimentally, the protocol quickly becomes complex and
demanding in terms of pulse design and frequency stability, above the second
order.

Our protocol uses two coherent microwave pulses: a main pulse drives the qubit
Rabi precession while a low-power, circularly polarized (image) pulse
continuously sustains the spin motion. The image drive has a frequency close to
the main drive and its amplitude is 1-2 orders of magnitude smaller. In this way, a
quantum gate could be driven by regular pulses, without the image pulse, while the time interval between gates could be filled with an integer number of Rabi nutations that use our protection protocol. Such scheme would protect the coherence of the qubit in-between usual quantum gates. We note that the initial phase difference between the two pulses allows to tune the spin dynamics by
enhancing (or diminishing) the Floquet modes\cite{Russomanno2017} of its second
dressing. The technical implementation is simple and can be generalized to any
type of qubit, such as superconducting circuits or spin systems. In this paper we focus on the experimental implementation and simply observe that numerical simulations based on Bloch model with $T_2=2T_1$ describe the final results very well.

\section{Results and discussion}

\begin{figure*}[t]
	\centering
	\includegraphics[width=0.95\textwidth]{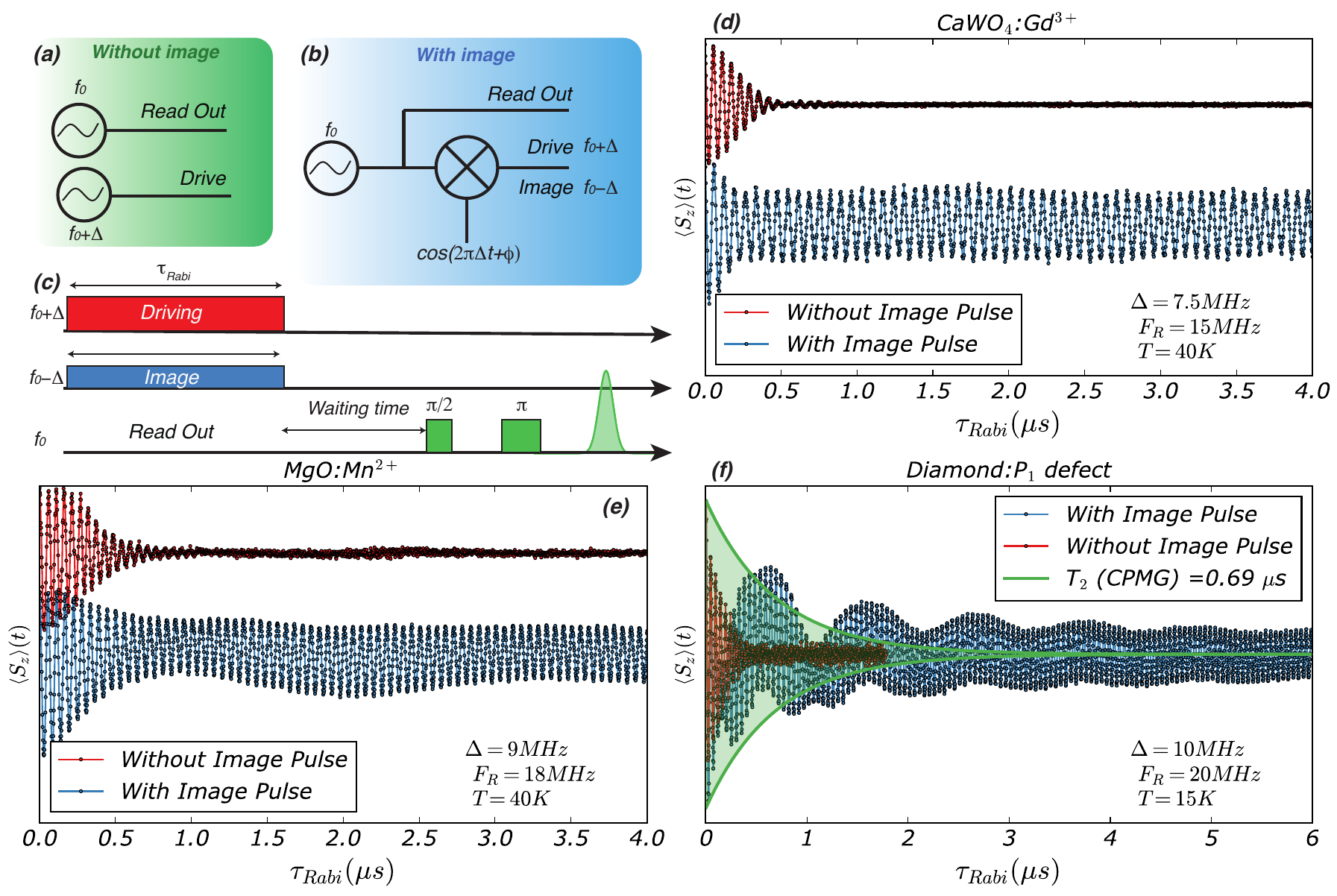}
	\caption{\TB{Comparison of Rabi oscillation with or without the image pulse.}
		\TB{a,b,c}, Schematic of the microwave implementation. \TB{a}, The drive
		($f+\Delta$)  and the readout ($f_0$) sources are independent. \TB{b}, Using
		the same source as the readout ($f_0$), the drive pulse ($f_0-\Delta$) is
		generated by a non-linear mixing with a low frequency signal $\Delta$, with
		$f_0/\Delta\sim10^3$, a process that creates a low-amplitude image drive
		($f_0-\Delta$) as well. The drive, image and readout pulses are coherent and
		the phase relationship is tunable. \TB{c} Pulse sequence (see \cite{Note19}):
		the drive and the image pulses act at the same time on the qubit while the
		readout is sensitive to the $S_z$ projection by a spin-echo process. \TB{d,e,f}
		Rabi oscillation of the \Gd, \Mn and P1 defects respectively in the presence
		(blue) or absence (red) of the image pulse at the optimum condition
		$F_R=2\Delta$. The green guideline in \TB{f} shows the improvement of the
		coherence time when compared to a CPMG pulse sequence while the blue curve
		shows not only a long coherence but also beatings which are tunable via
		phase-tunable Floquet dynamics (see text).
	}
	\label{Fig:compar}
\end{figure*}

The standard method to induce Rabi oscillations in a two-level system (TLS) is
to apply an electro-magnetic pulse of frequency $f_0$ equal to the TLS level
separation (resonance regime, $g\mu_B H_0=hf_0$). The pulse will drive the spin population
coherently between the two states. Experimentally, the drive is at a frequency
$f_0+\Delta$ (where $f_0$ is the Larmor frequency and $\Delta$ is a small
detuning away from the resonance condition) followed by read out pulses of
frequency $f_0$ to record the state $\mean{S_z}$ (see Fig.
\ref{Fig:compar}\TB{a}). The method introduced here makes use of two coherent
microwave pulses (see Fig. \ref{Fig:compar}\TB{b,c}): the drive pulse at $f_0+\Delta$ of amplitude $h_d$ and length $\tau_{Rabi}$
creates quantum Rabi oscillations while the second one sustain them using a very
low power image of the drive ($h_i\ll h_d$), operated at $f_0-\Delta$. In order to probe $\mean{S_z(\tau_{Rabi})}$ at the end of the Rabi sequence, we wait a time longer than $T_2$, such that
$\mean{S_x}\approx\mean{S_y}\approx0$, followed by $\pi/2- \pi$ pulses to create
a Hahn echo of intensity proportional to $\mean{S_z}$ (Mn$^{2+}$ spins are readout without echo, as detailed in Sect. IIB of \cite{Note19}).
Fig.~\ref{Fig:compar}\TB{b} shows one way of coherently creating the drive and
its image, by means of a mixer multiplying a pulse at frequency $f_0$ with an
intermediate frequency (IF) cosine signal allowing to control the detuning
$\Delta$, phase $\phi$ and the pulse length, shape and amplitude\footnote[19]{See Supplemental Material online for T2 characterization (Section I) and spin Hamiltonians (Section II) for the three materials, calculation of the rotating frame Hamiltonian and torque considerations (Section III), quantum protection for different initial states (Section IV) and power calibration of the setup (Section V).}. 

Rabi oscillations of three different types of paramagnetic systems $-$ a rare
earth ion (Gd$^{3+}$), a transition metal ion (Mn$^{2+}$) and a defect in
diamond (P1) $-$ are shown in Fig.~\ref{Fig:compar} \TB{(d,e,f)}, respectively.
Their detuned Rabi oscillations induced by the drive pulse only (red curves) are
of similar frequency ($\approx20$~MHz) and last for a small number of nutations
($<20$). The blue curve shows the Rabi oscillation when the image pulse is superimposed. The oscillations remain intense far
beyond the decay time of the red curves and their number is dramatically
increased. This effect has maximum impact when the frequency difference between
the drive and the image pulses $2\Delta$ matches the Rabi frequency induced by
the main drive, $F_R$. In addition of the very long coherence time, we observe a
slow amplitude modulation, depending on the phase $\phi$ and attributed to
Floquet modes, as explained below.

\begin{figure}[t]
	\centering
	\includegraphics[width=0.48\textwidth]{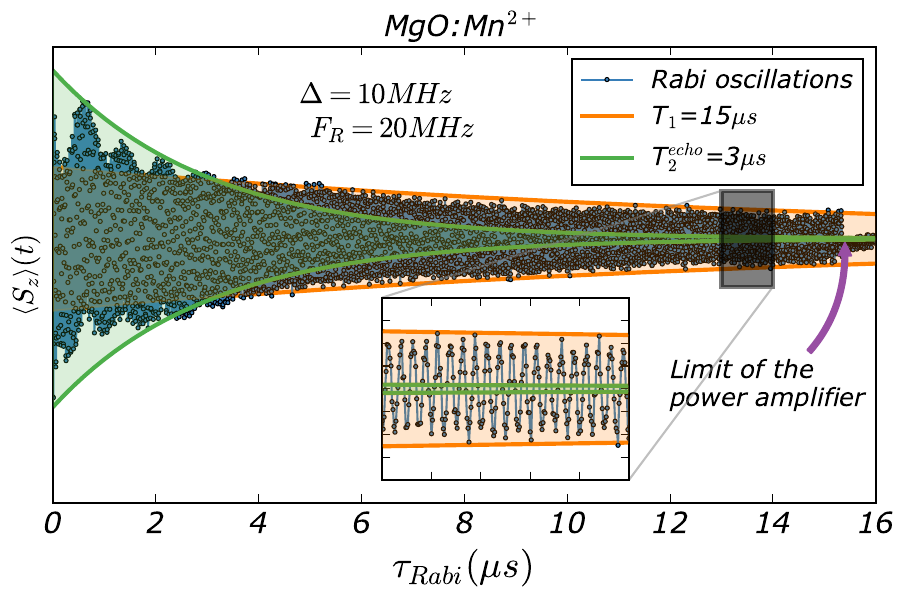}
	\caption{\textbf{Rabi decay under image pumping condition at 40~K}. The decay
		of the Rabi oscillations (green) for a Mn$^{2+}$ spin is limited by the
		relaxation time $T_1$ (orange) when an image pulse is used to sustain the
		dynamics. The inset shows the long-lived oscillations, all the way up to the
		15~$\mu$s limit as imposed by the power amplifier of the setup.}
	\label{Fig:T1T2}
\end{figure}
\begin{figure*}[t]
	\centering
	\includegraphics[width=0.95\textwidth]{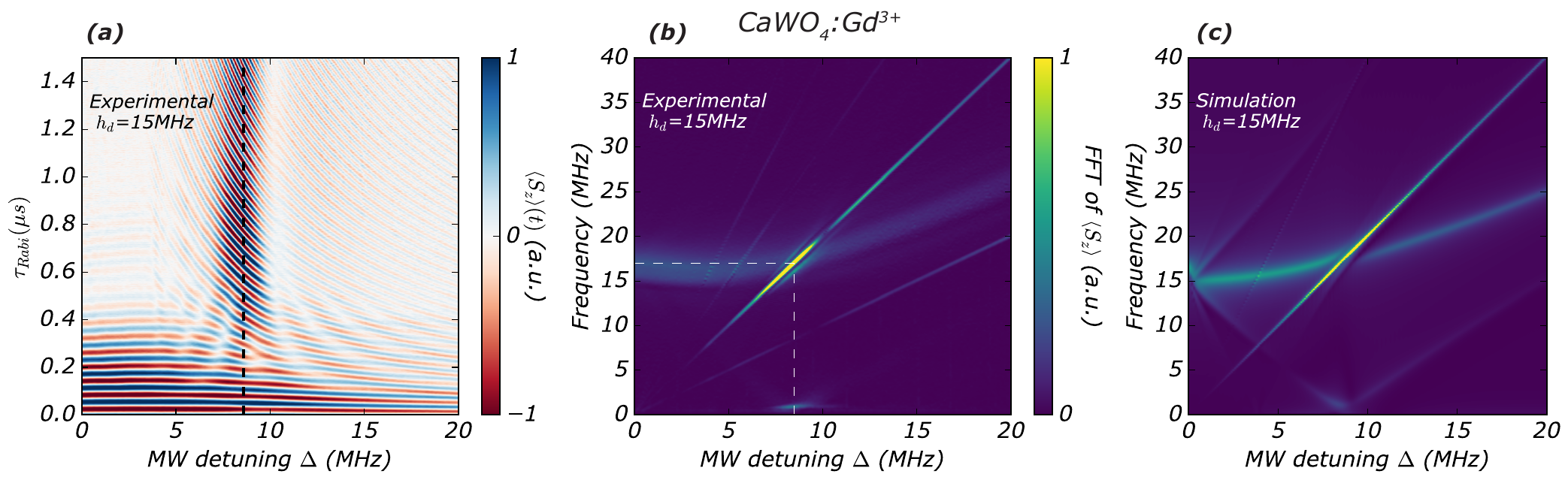}
	\caption{\textbf{Rabi dynamics as a function of detuning at fixed drive level
			$h_d$}. \TB{a}, Contour plot showing $S_z$ as a function of pulse length
		$\tau_{Rabi}$ and microwave detuning $\Delta$. At $F_R=2\Delta$ (dashed), the
		Rabi oscillations are optimally sustained by the image pulse. \TB{b}, The FFT
		of \TB{a} shows an intense peak where the optimum protection is achieved (dashed lines).
		\TB{c}, A simulated FFT using the rotating wave approximation and the
		parameters of \TB{b}, is in agreement with the experimental observations.}
	\label{Fig:detuning}
\end{figure*}
We study the new decay time of the Rabi oscillation under image pumping, by tuning the relaxation time via temperature control and by applying
the longest drive pulse available to us (Fig. \ref{Fig:T1T2}). The length of the drive pulse is limited by the pulse power amplifier of the setup, with a maximum
pulse length of 15~$\mu$s. At 40~K, the relaxation time of the spin system
MgO:Mn$^{2+}$ ($S=5/2$) is also $\approx15\mu$s. The Rabi oscillation for the
transition $+\frac{1}{2}\leftrightarrow -\frac{1}{2}$ is shown in
Fig.~\ref{Fig:T1T2}. Guidelines showing the exponential decays due to $T_2$ (in
green) and $T_1$ (in orange) measured by Hahn echo and inversion recovery
respectively, are added as well. While the amplitude of the Floquet mode (slow
amplitude modulation) decreases with $T_2$, the Rabi oscillations persists with
a decay time $\approx T_1$.

\subsection*{Protection of quantum coherence for different initial states}  

We analized our protocol for different initial states. For initial and final states along $+X$, $+Y$ and $+Z$ we observe that the protection protocol is effective for times up to the maximum amplifier gate length of 15~$\mu$sec. 

\paragraph{Pulse sequence}

\begin{figure}[h]
	\centering
	\includegraphics[width=0.9\textwidth]{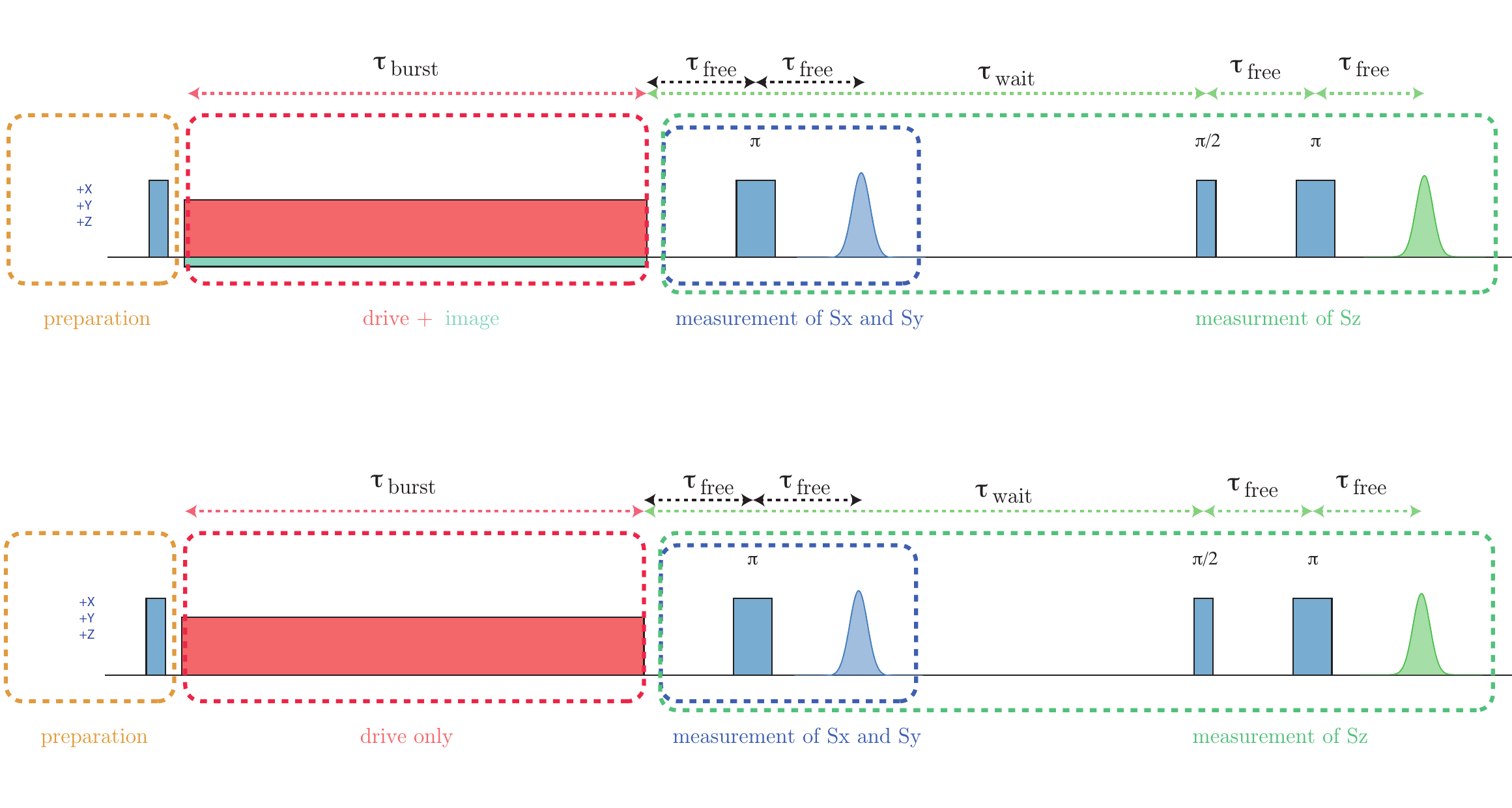}
	\caption{\textbf{Pulse sequence used for different initial states.} First, the spin is prepared in the ground state $+Z$ or, after a $(\frac{\pi}{2})_{y,x}$, in the $+X$ or $+Y$ states. A Rabi oscillation is performed during $\tau_{burst}$ for an integer number of Rabi flops, with or without the image pulse $h_i$ (top and bottom panels respectively). The final state is along the same direction as the initial one, and the readout is done using an usual echo detection (blue and green areas). }
	\label{FigSI:sequence}
\end{figure}

The protocol proposed here was tested for different initial states and spin systems. In the following, the details of the experiment are given. The ground state of the spin is along $+Z$ axis and is used as initial state. In Fig.~\ref{FigSI:sequence}, this is shown in orange and labeled ``preparation". The $+Z$ state is obtained by thermalization. States $+X$ and $+Y$ are obtained from $+Z$ using a hard $\frac{\pi}{2}$ pulse around the $y$ or $x$ axes, respectively, able to excite the whole spin ensemble (or spectroscopic line). In \Mn and P1, this is ensured by their narrow spectroscopic linewidths $\Gamma$, while in \Gd , $\Gamma$ is of the same order of magnitude with the maximum excitation bandwidth. Therefore, some Gd$^{3+}$ spins of the spin packet might not be in a perfect $+X$ or $+Y$ state. However we didn't notice any particular effect in the final results for Gd. We note that by combining rotations around any of the $x$, $y$ and $z$ axes, we can prepare the initial state in any position. Once prepared, the spin state is subjected to coherence protected Rabi protocol or to the usual Rabi drive for an integer number of Rabi flops (top and bottom ``burst" panels in Fig.~\ref{FigSI:sequence}, respectively). Thus, the final state is along the same direction as the initial one \cite{Morton2008}. The echo-based measurement for an initial state $+Z$ is shown in the green panel (with $\tau_{wait}\gg T_2$). For initial states $+X$ and $+Y$ (blue panel), one have to wait a time $\tau_{free}$ in order to let the spin packet defocus before applying a $\pi$ pulse and detect the subsequent echo signal.

Throughout this article, the $\pi/2$ and $\pi$ pulse lengths are 14 and 28 ns, respectively. The readout waiting times $\tau_{wait}$ are approximately 5~$\mu$s for P1 defects, 6~$\mu$s for Mn$^{2+}$ and 10~$\mu$s for Gd$^{3+}$.

\paragraph{P1 defects in diamond}

\begin{figure}[h]
	\centering
	\includegraphics[width=0.45\textwidth]{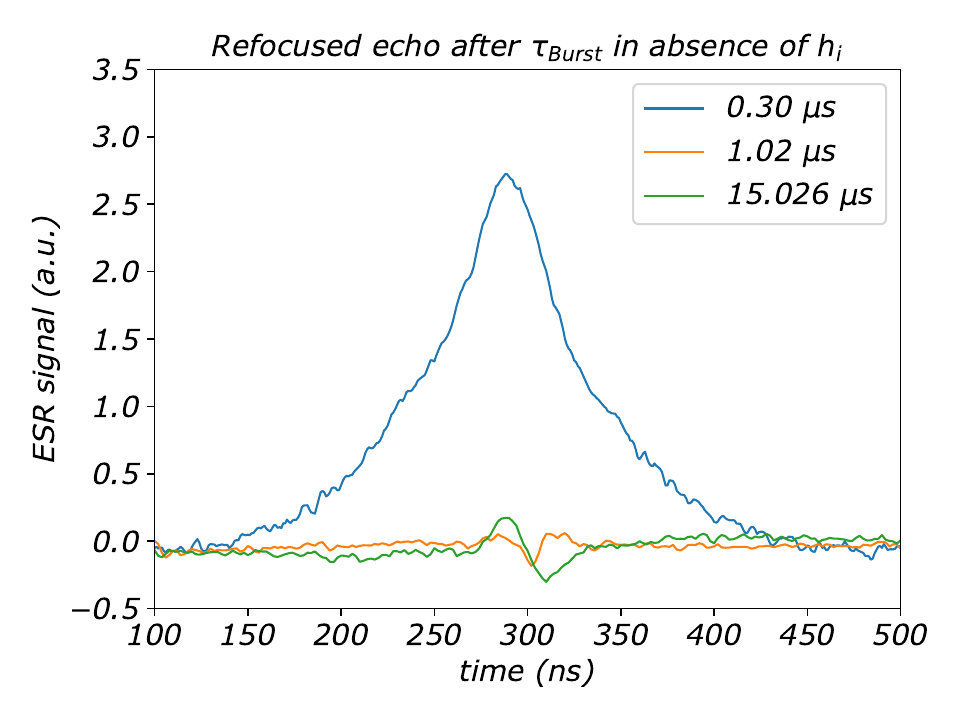}
	\includegraphics[width=0.45\textwidth]{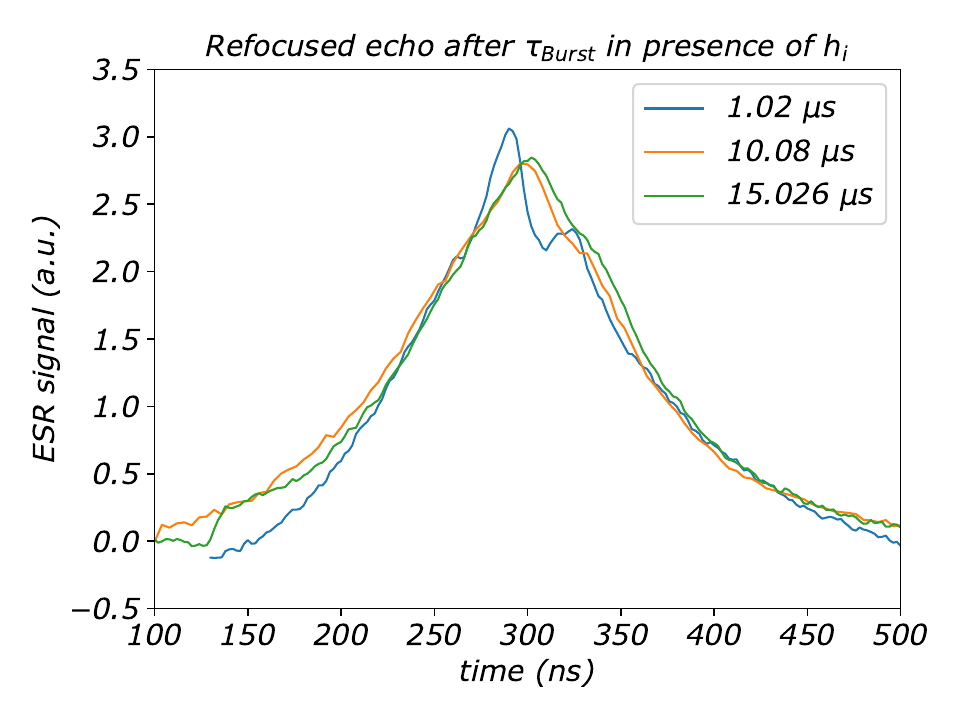}
	\caption{\textbf{P1 defects in diamond}. Echo signal for initial and final state along $+Z$ measured after a Rabi burst of various length. Without image pulse (left panel) the coherence is quickly lost after 0.30~$\mu$s as shown by an absence of signal above 1~$\mu$s. With protection protocol in place (right panel), the signal is well conserved up to maximum burst length of 15~$\mu$s.}
	\label{FigSI:BenchP1}
\end{figure}

In the case of P1 defects in diamond, we present measurements of the spin echo signal when the initial and final states are along $+Z$, with and without the protection protocol for different lengths of the Rabi pulse $\tau_{burst}$. Without image pulse (Fig.~\ref{FigSI:BenchP1} left) the signal is visible after a Rabi burst of 300 ns, but it is rapidly lost for times longer than 1~$\mu$s. With the protection protocol in place (Fig.~\ref{FigSI:BenchP1} right), the signal is almost entirely conserved for times up to 15~$\mu$s. This behavior shows a significant improvement over the CPMG method (see \emph{Supplementary Information}, $T_2=0.69$~$\mu$s). For this experiment the temperature was set to $T=$15 K and $\tau_{free}=300$~ns. 

\paragraph{CaWO$_4$:Gd$^{3+}$}

\begin{figure}[h]
	\centering
	\includegraphics[width=0.9\textwidth]{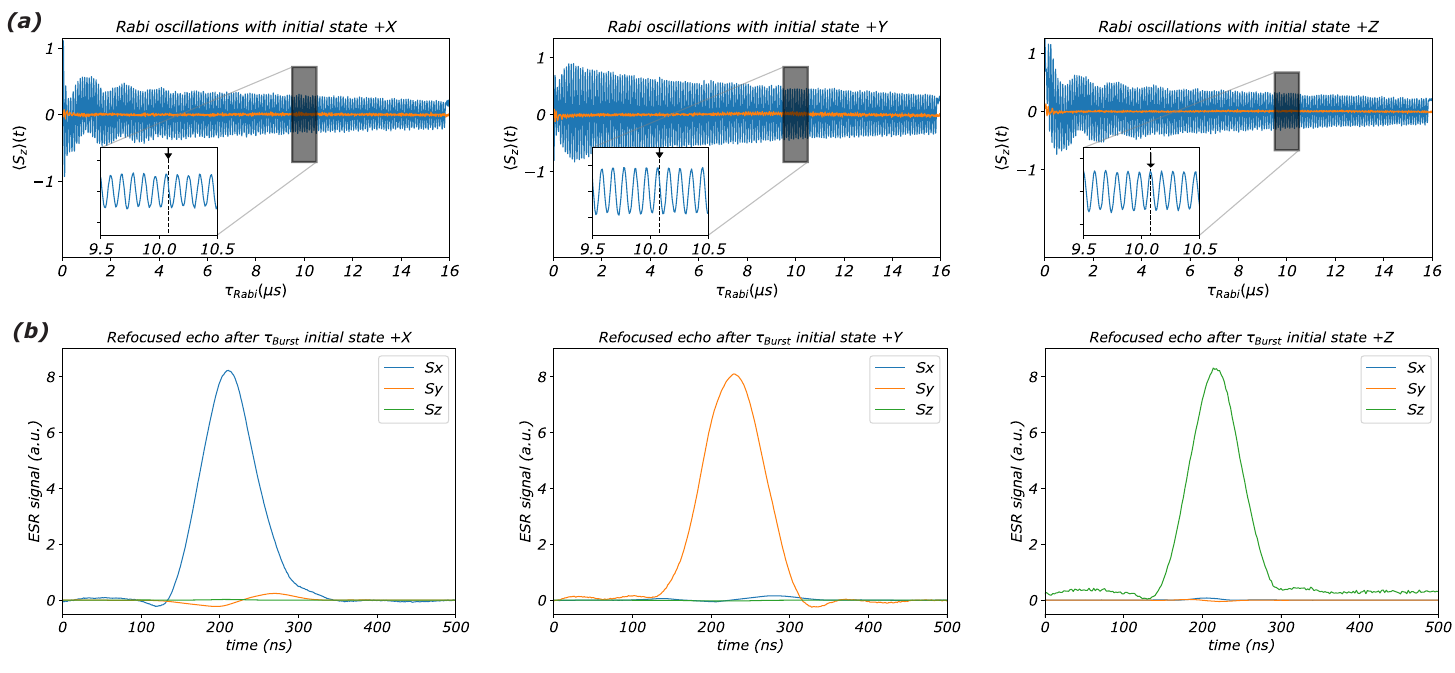}
	\caption{\textbf{CaWO$_4$:Gd$^{3+}$}. (a) Rabi oscillations shown as the real and imaginary part of the recorded signal (blue and orange lines, respectively). From left to right, the initial (and final) state of the spin are $+X$, $+Y$ and $+Z$. After a time given by an integer number of Rabi flops, shown with an arrow in the inset, (b) the spin-echo signal is recorded for each spin component, using the measurement sequence described in Fig.~\ref{FigSI:sequence}. }
	\label{FigSI:benchmark}
\end{figure}

For Gd$^{3+}$ spins, we measured the spin-echo signal after an integer number of Rabi flops for initial/final states along the $+X$, $+Y$ and $+Z$ axes. The Rabi oscillations are shown in Fig.~\ref{FigSI:benchmark} (a) as the real and imaginary part of the recorded signal (blue and orange lines, respectively). One notes that for the initial state along $+X$ and $+Y$ the Rabi signal starts and ends at zero value, with the end time being indicated by the arrow in each inset. The bottom panel shows large echo signals after 10~$\mu$s, much longer than $T_{2CPMG}=4$~$\mu$s (see \emph{Supplementary Information}), for each of the initial state preparation. Using a combination of rotations around the $x$,$y$ and $z$ axes, we can create any initial state for the spin and thus use the image pulse protection for any arbitrary state. The experimental conditions for these measurements are: temperature $T=$40~K and $\tau_{free}=200$~ns.

\subsection*{Qubit dynamics}  

The qubit dynamics in the absence of a bath, is described by
the spin Hamiltonian in the laboratory frame \cite{Note19} :
\begin{equation}\label{eq:HamLabFram} \mathcal H=
f_0S_z+2h_{d}S_x\sin(\omega_+t+\phi)+2h_{i}S_x\sin(\omega_-t-\phi-\theta)
\end{equation} where $f_0$ is the Larmor frequency caused by the static field,
$h_d$ and $h_i$ are the microwave drive and image field, respectively,
$\frac{\omega_{+,-}}{2\pi}=f_0\pm\Delta$, $\phi$ is a tunable phase (see
Fig.~\ref{Fig:compar}) and $\theta$ is a small additional phase, possibly
created by imperfections of the setup (as discussed in \cite{Note19} ). Variables $f_0,h_{d,i}$ and $\Delta$ are expressed in units of
MHz. After a transformation in a frame rotating with $\omega_+$, the Hamiltonian
\eqref{eq:HamLabFram} becomes : \begin{align}\label{eq:HamRotFram} &\mathcal
H_{RF}=-\Delta S_z+h_{d}(S_x\sin\phi-S_y\cos\phi)\\ &-h_{i}[S_x\sin(4\pi\Delta
t+\phi+\theta)+S_y\cos(4\pi\Delta t+\phi+\theta)]. \nonumber \end{align} When
the image field $h_i$ is absent, the equation \eqref{eq:HamRotFram} has no
explicit time dependence and  the Rabi frequency is simply
$F_R=\sqrt{\Delta^2+h_d^2}$. When $h_i$ is present, the dynamics of $\mean{S_z}$
can be solved numerically, as it is shown in Fig.~\ref{Fig:detuning} for the
case of \Gd.

For a fixed power $h_d$ of the drive pulse, Rabi oscillations are measured as a
function of the detuning $\Delta$. As shown in the contour plot of
Fig.~\ref{Fig:detuning}\TB{a}, $\mean{S_z}(t)$ vanishes after few oscillations
except when the condition $F_R\sim2\Delta$ is met. At this Floquet resonance,
$\mean{S_z}$ keeps oscillating for a very long time ($>15\mu$s). Its Fast
Fourier Transform (FFT) is presented in Fig.~\ref{Fig:detuning}\TB{b}. The free
(unprotected) Rabi oscillations mode $F_R$ is rather weak and broad showing the
large damping caused by the environment. However, when the mode crosses the
frequency of the image pulse, indicated by the vertical white dashed line, the
peak becomes intense and narrower, as the qubit protection from environment
is activated. The condition  $F_R=\sqrt{\Delta^2+h_d^2}=2\Delta$ (or
$h_d=\sqrt{3}\Delta$) gives the most efficient protection of the Rabi
oscillation (see \cite{Note19}). The general condition is $F_R=n\Delta,n=2k, k\in N$ showing a comensurate motion of the qubit and $h_i$ on the Bloch sphere.

We can compare the experimental result to the model described by the Hamiltonian
\eqref{eq:HamRotFram} which can be rewritten as in Eq.~S14 in \cite{Note19} : $\mathcal H_{RF}=-\Delta
S_z+S_+[h_{d}e^{-i(\phi-\pi/2)}+h_{i}e^{i(4\pi\Delta t+\phi+\theta+\pi/2)}]$.
When the "image" pulse is not applied, the Hamiltonian is time independent and
the propagator is simply the matrix exponential of the Hamiltonian:
$U_p(t)=\exp(-i2\pi\mathcal H_{RF} t)$. When the image pulse is present
($h_i>0$), the Hamiltonian becomes explicitly time-dependent. Although a second
canonical transformation RWA could remove the time dependence if $\Delta\gg
h_i$, it is importat to leave $\Delta$ as a free parameter since the methods
works at resonance as well ($\Delta=0$). Thus, for the sake of generality, we
solved numerically the explicit time-dependent differential equations using \texttt{QuTIP }  \cite{Johansson2013}. The parameters used in
the simulation have been measured independently: the microwave drive field $h_d$
has been calibrated using the frequency of Rabi oscillations at no detuning
($\Delta=0$), the image drive $h_i$ was measured by a spectrum analyzer directly
connected to the output of the AWG ($h_i/h_d\approx 0.12$), relaxation ($T_1$) and decoherence ($T_2$)
times were measured by inversion recovery and Carr-Purcell-Meiboom-Gill (CPMG)
protocol, respectively (see \cite{Note19}). We used \texttt{QuTiP} implementation of Lindblad's master equation with $S_-$ as collapse operator which is equivalent to the phenomenological Bloch model for the case $T_2=2T_1$. Figure \ref{Fig:detuning}\TB{c} shows the FFT of $\mean{S_z}(t)$ computed using the time evolution of $\mathcal H_{RF}$. The Hamiltonian
\eqref{eq:HamRotFram} describes very well the protection of the coherence by
means of the image pulse. Note the existence of a Floquet mode at
$\Delta$=7.5~MHz of frequency $\sim1$~MHz, visible in both the experimental and
theoretical contour plots of Fig.~\ref{Fig:detuning}.

\begin{figure}[t] \centering
	\includegraphics[width=0.48\textwidth]{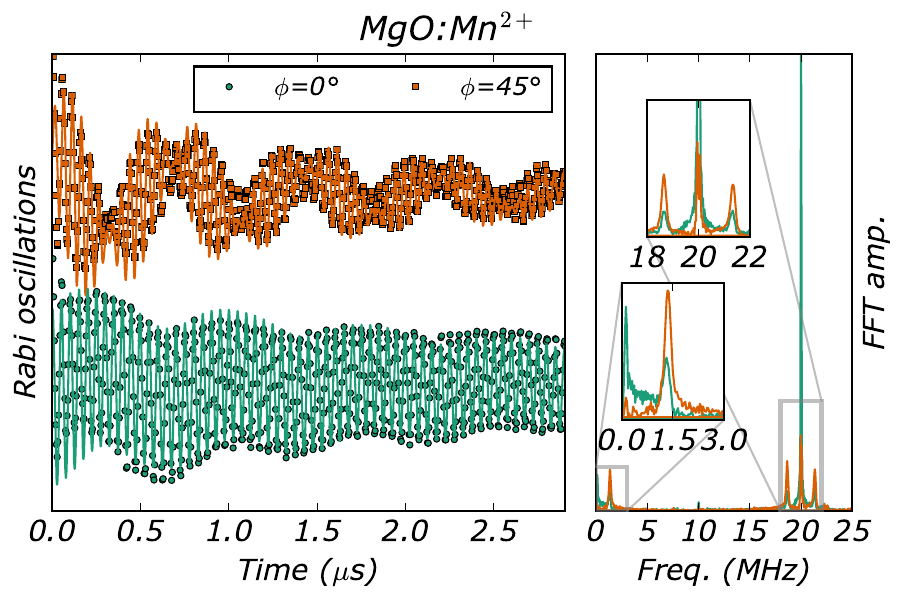}
	\caption{\textbf{Phase-tunable Floquet dynamics}. Rabi oscillations of Mn$^{2+}$
		at 40~K for two values of $\phi$. For $\phi=0^\circ$ (green symbols), $h_d$ and $h_i$ pulses have the same initial phases resulting in optimally
		sustained Rabi oscillation. For $\phi=45^\circ$ (orange symbols), strong beatings are observed
		($h_{d,i}$ terms of $\mathcal H_{RF}$ are $\perp$). The solid lines are
		simulations using our model with no fitting parameters. The right panel shows the
		experimental FFT traces. The green one has a well-defined peak at $F_R$
		while in the other one two peaks (top inset) are separated by twice the
		Floquet mode frequency (bottom inset).  } \label{Fig:MgOPhase} \end{figure}

The Floquet mode appears as beatings of the Rabi frequency and is
$\phi$-tunable. Similarly to the case of Gd$^{3+}$, the qubit protection and
Floquet mode dynamics is obtained for the $S=5/2$ spin of \Mn, here measured under
the experimental conditions of Fig.~\ref{Fig:T1T2}. Rabi oscillations and
corresponding FFT spectra are shown in Fig.~\ref{Fig:MgOPhase} for two values of
the initial phase, $\phi=0^\circ$ (green) and $\phi=45^\circ$ (gold), while
simulations are shown in black. The decay times are much larger than $T_2$ for
both values (here $T_1\approx15$~$\mu$s, see Fig.~\ref{Fig:T1T2}); however, the
dynamics is strikingly different. When the drive and image pulses have the same
initial phase (one can consider the initial time in Eq.~\ref{eq:HamRotFram} as
$-\frac{\theta}{4\pi\Delta}$ without loss of generality), the Rabi oscillations
have maximum visibility, with almost no beatings (see \cite{Note19}). At $\phi=45^\circ$, the spin torques
generated by the $h_d$ and $h_i$ fields induce strong beatings or a Floquet mode
creating two additional modes of the Rabi frequency. The left panel shows Rabi
splittings equal to the Floquet frequency for $\phi=45^\circ$ and a single Rabi
oscillation for $\phi=0^\circ$.

\begin{figure}[t] \centering
	\includegraphics[width=0.48\textwidth]{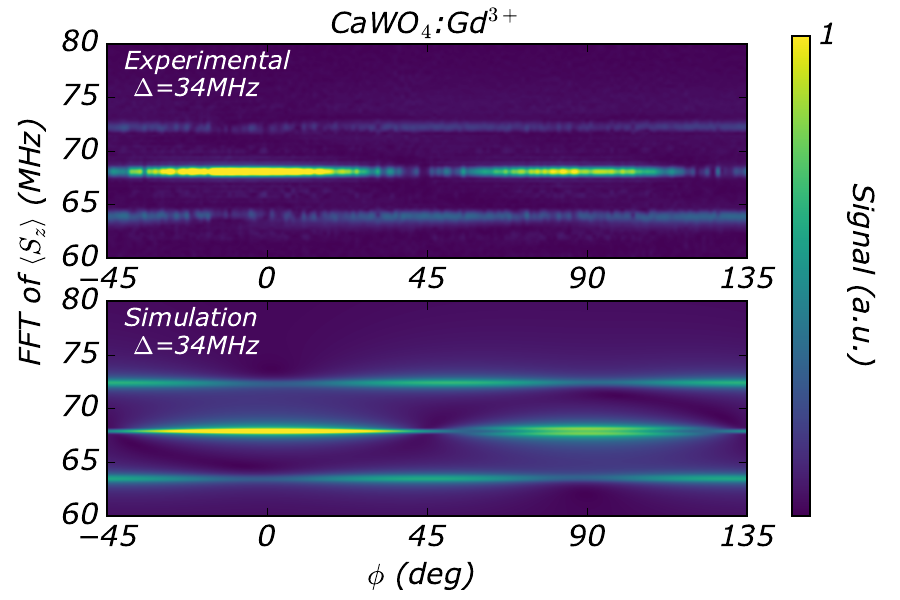}
	\caption{\textbf{Phase-dependence of the Floquet mode}. Experimental (top) and
		simulated (bottom) FFT of Rabi oscillations for $\Delta=h_d/\sqrt3=34$~MHz as a
		function of $\phi$ (amplitude is normalized to the highest peak). For
		$\phi=2k\pi/4, k\in N$, Rabi oscillations have a single mode $F_R$ while for
		$\phi=(2k+1)\pi/4$ a splitting of twice the Floquet mode frequency is observed.
		While its frequency is fixed, the intensity of the Floquet mode changes
		gradually, with a period of $\pi/2$.} \label{Fig:Phase2D} \end{figure}

Experimentally, we can continuously vary the value of $\phi$ and analyze the
frequency and intensity of the Floquet mode. As an example, a comparison between
theory and experiment is shown in Fig.~\ref{Fig:Phase2D} for the case of \Gd 
for $\Delta=h_d/\sqrt3=34$~MHz. For even and odd multiples of $\pi/4$, a single
and a splitted Rabi mode is observed, respectively. The Rabi splitting is the
Floquet mode and is constant as a function of $\phi$ but its intensity
oscillates with a period of $\pi/2$. The effect is evident in simulations as well, since the terms in $h_{d,i}$ of $\mathcal H_{RF}$ are along the same direction or orthogonal, for $\phi=0^\circ$ and 45$^\circ$ respectively.

In regards to decoherence sources, it is safe to assume that the main contribution comes from the spin bath made by nuclear spins surrounding the central spin, as well as other electronic spins located in its closed vicinity. Such scenario is the typical situation in spin systems operated at low enough temperatures to reduce the role of the phonon bath on $T_2$. The details of the entangled qubit-bath dynamics are outside the scope of the current study. We do observe that the final results are well described when only dissipation is the source of decoherence, leading to $T_2=2T_1$. This may indicate that the image pulse $h_i$ is able to control the dynamics and thus the decoherence of the spin bath.

The qubit rotation is thus tunable by using a pre-selected value
of $\phi$, allowing to create complex rotations. With a decoherence time
approaching spin lifetime $T_1$, the value of $\phi$ can be changed while qubit
control is still ongoing. Our study demonstrates a sustained quantum coherence
using a general protocol that can be readily implemented to
any type of qubit. Our approach can be used in other detection
schemes, such as sensitive spin detection using on-chip resonance
techniques\cite{Yue2017,Probst2017,Toida2019}.

\section*{Methods}

\TI{Spectrometer setup.} The measurements have been performed on a conventional pulse ESR spectrometer Bruker E680 equipped with an incoherent electron double resonance (ELDOR) bridge and a coherent arbitrary waveform generator (AWG) bridge. In the ELDOR bridge (Fig. \ref{Fig:compar}\TB{a}), the drive and the read out pulses come from two independent sources while with the AWG bridge (Fig.\ref{Fig:compar}\TB{b}) all the pulses are generated using the same microwave source and thus they are all phase coherent.  The drive frequency is generated by mixing the source $f_0$ (used as a local oscillator) with a low frequency and phase controllable signal $\DP$ through an in-phase quadrature (IQ) mixer. Ideally, the output of the mixer is monochromatic with the frequency $f_0+\Delta$. In reality, the output consists of a principal frequency $f_0+\Delta$ (the drive) and of lower amplitude images $f_0+n\Delta$ (see \TI{Supplementary Information} for more information). Since the effect of the image is the central part of this paper, we have characterized the AWG bridge using a spectrum analyzer, right before the power amplification stage. An example of spectrum is presented in Fig.~S6 of the \TI{Supplementary Information}. The power of the image $f_0-\Delta$ is lower by $\approx -18$~dB than $f_0+\Delta$. Consequently, an amplitude ratio of the MW magnetic fields $h_i/h_d$ around $\sim$0.12 is used in simulations.

\TI{Pulse sequence}. First, the system is set to be in resonance condition $g\mu_B H_0=hf_0$. The drive pulse of amplitude $h_D$, frequency $f_0+\Delta$  and length $\tau_{Rabi}$ induces Rabi oscillation in detuning regime. At the same moment the image pulse (generated through the IQ mixer) of amplitude $h_I$, frequency $f_0-\Delta$ and the same length $\tau_{Rabi}$ also irradiate the spins. In order to probe $\mean{S_z(\tau_{Rabi})}$ at the end of the Rabi sequence, we wait a time longer than $T_2$, such that $\mean{S_x}\approx\mean{S_y}\approx0$, followed by $\pi/2- \pi$ pulses to create a Hahn echo of intensity proportional to $\mean{S_z}$.

\TI{Spin systems}. The methodology presented here is demonstrated on different spin systems: the nitrogen substitution in diamond P1 defect ($S=1/2$) (concentration :100~ppm), Mn$^{2+}$ impurities in MgO ($S=5/2$) with a concentration of 10~ppm and Gd$^{3+}$ impurities in CaWO$_4$ ($S=7/2$) with a concentration of 50~ppm.  Despite the large Hilbert space of the Mn$^{2+}$ and Gd$^{3+}$ spin Hamiltonians, the orientation of the magnetic field and the frequency and power of the microwave excitation are chosen to avoid multiple level transitions and thus select only one resonance \cite{Bertaina2009}. Therefore, the spin systems can be considered as effective two-level systems undergoing coherent Rabi rotations. The spin Hamiltonians, operating parameters (fields and frequencies) as well as characteristic $T_{1,2}$ times for these materials are given in the Supplementary Information.

\TI{Simulation explanation}. In the rotating frame, the Hamiltonian describing the dynamics of the electron spin is given by Eq.~S14 in the \TI{Supplementary Information}: $\mathcal H_{RF}=-\Delta S_z+S_+[h_{d}e^{-i(\phi-\pi/2)}+h_{i}e^{i(4\pi\Delta t+\phi+\theta+\pi/2)}]$. When the "image" pulse is not applied, the Hamiltonian is time independent and the propagator is simply the matrix exponential of the Hamiltonian: $U_p(t)=\exp(-i2\pi\mathcal H_{RF} t)$. When the image pulse is present ($h_i>0$), the Hamiltonian becomes explicitly time-dependent. Although a second canonical transformation RWA could remove the time dependence if $\Delta\gg h_i$, it is importat to leave $\Delta$ as a free parameter since the methods works at resonance as well ($\Delta=0$). Thus, for the sake of generality, we solved numerically the explicit time-dependent differential equations using the quantum toolbox \texttt{QuTIP }  \cite{Johansson2013}. The parameters used in the simulation have been measured independently: the microwave drive field $h_d$ has been calibrated using the frequency of Rabi oscillations at no detuning ($\Delta=0$), the image drive $h_i$ was measured by a spectrum analyzer directly connected to the output of the AWG, relaxation ($T_1$) and decoherence ($T_2$) times were measured by inversion recovery and Carr-Purcell-Meiboom-Gill (CPMG) protocol, respectively (see \TI{Supplementary Information, Sect I}).

\bibliographystyle{naturemag.bst} 
\bibliography{Protect_MAIN.bib}

\begin{thebibliography}{10}
\expandafter\ifx\csname url\endcsname\relax
  \def\url#1{\texttt{#1}}\fi
\expandafter\ifx\csname urlprefix\endcsname\relax\def\urlprefix{URL }\fi
\providecommand{\bibinfo}[2]{#2}
\providecommand{\eprint}[2][]{\url{#2}}

\bibitem{Chirolli2008}
\bibinfo{author}{Chirolli, L.} \& \bibinfo{author}{Burkard, G.}
\newblock \bibinfo{title}{Decoherence in solid-state qubits}.
\newblock \emph{\bibinfo{journal}{Advances in Physics}}
  \textbf{\bibinfo{volume}{57}}, \bibinfo{pages}{225--285}
  (\bibinfo{year}{2008}).

\bibitem{Yoneda2018a}
\bibinfo{author}{Yoneda, J.} \emph{et~al.}
\newblock \bibinfo{title}{A quantum-dot spin qubit with coherence limited by
  charge noise and fidelity higher than 99.9\%}.
\newblock \emph{\bibinfo{journal}{Nature Nanotechnology}}
  \textbf{\bibinfo{volume}{13}}, \bibinfo{pages}{102--106}
  (\bibinfo{year}{2018}).

\bibitem{Viola1998}
\bibinfo{author}{Viola, L.} \& \bibinfo{author}{Lloyd, S.}
\newblock \bibinfo{title}{Dynamical suppression of decoherence in two-state
  quantum systems}.
\newblock \emph{\bibinfo{journal}{Physical Review A}}
  \textbf{\bibinfo{volume}{58}}, \bibinfo{pages}{2733--2744}
  (\bibinfo{year}{1998}).

\bibitem{Viola1999}
\bibinfo{author}{Viola, L.}, \bibinfo{author}{Knill, E.} \&
  \bibinfo{author}{Lloyd, S.}
\newblock \bibinfo{title}{Dynamical {{Decoupling}} of {{Open Quantum
  Systems}}}.
\newblock \emph{\bibinfo{journal}{Physical Review Letters}}
  \textbf{\bibinfo{volume}{82}}, \bibinfo{pages}{2417--2421}
  (\bibinfo{year}{1999}).

\bibitem{Cai2012}
\bibinfo{author}{Cai, J.-M.} \emph{et~al.}
\newblock \bibinfo{title}{Robust dynamical decoupling with concatenated
  continuous driving}.
\newblock \emph{\bibinfo{journal}{New Journal of Physics}}
  \textbf{\bibinfo{volume}{14}}, \bibinfo{pages}{113023}
  (\bibinfo{year}{2012}).

\bibitem{Cohen2017}
\bibinfo{author}{Cohen, I.}, \bibinfo{author}{Aharon, N.} \&
  \bibinfo{author}{Retzker, A.}
\newblock \bibinfo{title}{Continuous dynamical decoupling utilizing
  time-dependent detuning}.
\newblock \emph{\bibinfo{journal}{Fortschritte der Physik}}
  \textbf{\bibinfo{volume}{65}}, \bibinfo{pages}{1600071}
  (\bibinfo{year}{2017}).

\bibitem{Farfurnik2017}
\bibinfo{author}{Farfurnik, D.} \emph{et~al.}
\newblock \bibinfo{title}{Experimental realization of time-dependent
  phase-modulated continuous dynamical decoupling}.
\newblock \emph{\bibinfo{journal}{Physical Review A}}
  \textbf{\bibinfo{volume}{96}}, \bibinfo{pages}{013850}
  (\bibinfo{year}{2017}).

\bibitem{Teissier2017}
\bibinfo{author}{Teissier, J.}, \bibinfo{author}{Barfuss, A.} \&
  \bibinfo{author}{Maletinsky, P.}
\newblock \bibinfo{title}{Hybrid continuous dynamical decoupling: A
  photon-phonon doubly dressed spin}.
\newblock \emph{\bibinfo{journal}{Journal of Optics}}
  \textbf{\bibinfo{volume}{19}}, \bibinfo{pages}{044003}
  (\bibinfo{year}{2017}).

\bibitem{Rohr2014}
\bibinfo{author}{Rohr, S.} \emph{et~al.}
\newblock \bibinfo{title}{Synchronizing the {{Dynamics}} of a {{Single Nitrogen
  Vacancy Spin Qubit}} on a {{Parametrically Coupled Radio}}-{{Frequency
  Field}} through {{Microwave Dressing}}}.
\newblock \emph{\bibinfo{journal}{Physical Review Letters}}
  \textbf{\bibinfo{volume}{112}}, \bibinfo{pages}{010502}
  (\bibinfo{year}{2014}).

\bibitem{Welinski2019}
\bibinfo{author}{Welinski, S.} \emph{et~al.}
\newblock \bibinfo{title}{Electron {{Spin Coherence}} in {{Optically Excited
  States}} of {{Rare}}-{{Earth Ions}} for {{Microwave}} to {{Optical Quantum
  Transducers}}}.
\newblock \emph{\bibinfo{journal}{Physical Review Letters}}
  \textbf{\bibinfo{volume}{122}} (\bibinfo{year}{2019}).

\bibitem{Lim2018}
\bibinfo{author}{Lim, H.-J.}, \bibinfo{author}{Welinski, S.},
  \bibinfo{author}{Ferrier, A.}, \bibinfo{author}{Goldner, P.} \&
  \bibinfo{author}{Morton, J. J.~L.}
\newblock \bibinfo{title}{Coherent spin dynamics of ytterbium ions in yttrium
  orthosilicate}.
\newblock \emph{\bibinfo{journal}{Physical Review B}}
  \textbf{\bibinfo{volume}{97}}, \bibinfo{pages}{064409}
  (\bibinfo{year}{2018}).
\newblock \eprint{1712.00435}.

\bibitem{Bienfait2016}
\bibinfo{author}{Bienfait, A.} \emph{et~al.}
\newblock \bibinfo{title}{Controlling spin relaxation with a cavity}.
\newblock \emph{\bibinfo{journal}{Nature}} \textbf{\bibinfo{volume}{531}},
  \bibinfo{pages}{74--77} (\bibinfo{year}{2016}).

\bibitem{Facchi2004}
\bibinfo{author}{Facchi, P.}, \bibinfo{author}{Lidar, D.~A.} \&
  \bibinfo{author}{Pascazio, S.}
\newblock \bibinfo{title}{Unification of dynamical decoupling and the quantum
  {{Zeno}} effect}.
\newblock \emph{\bibinfo{journal}{Physical Review A}}
  \textbf{\bibinfo{volume}{69}}, \bibinfo{pages}{032314}
  (\bibinfo{year}{2004}).

\bibitem{Fanchini2007}
\bibinfo{author}{Fanchini, F.~F.}, \bibinfo{author}{Hornos, J. E.~M.} \&
  \bibinfo{author}{Napolitano, R. d.~J.}
\newblock \bibinfo{title}{Continuously decoupling single-qubit operations from
  a perturbing thermal bath of scalar bosons}.
\newblock \emph{\bibinfo{journal}{Physical Review A}}
  \textbf{\bibinfo{volume}{75}}, \bibinfo{pages}{022329}
  (\bibinfo{year}{2007}).

\bibitem{Shu2018}
\bibinfo{author}{Shu, Z.} \emph{et~al.}
\newblock \bibinfo{title}{Observation of {{Floquet Raman Transition}} in a
  {{Driven Solid}}-{{State Spin System}}}.
\newblock \emph{\bibinfo{journal}{Physical Review Letters}}
  \textbf{\bibinfo{volume}{121}}, \bibinfo{pages}{210501}
  (\bibinfo{year}{2018}).

\bibitem{Saiko2018}
\bibinfo{author}{Saiko, A.~P.}, \bibinfo{author}{Markevich, S.~A.} \&
  \bibinfo{author}{Fedaruk, R.}
\newblock \bibinfo{title}{Multiphoton {{Raman}} transitions and {{Rabi}}
  oscillations in driven spin systems}.
\newblock \emph{\bibinfo{journal}{Physical Review A}}
  \textbf{\bibinfo{volume}{98}}, \bibinfo{pages}{043814}
  (\bibinfo{year}{2018}).

\bibitem{Yu2020}
\bibinfo{author}{Yu, M.} \emph{et~al.}
\newblock \bibinfo{title}{Experimental measurement of the quantum geometric
  tensor using coupled qubits in diamond}.
\newblock \emph{\bibinfo{journal}{National Science Review}}
  \textbf{\bibinfo{volume}{7}}, \bibinfo{pages}{254--260}
  (\bibinfo{year}{2020}).

\bibitem{Russomanno2017}
\bibinfo{author}{Russomanno, A.} \& \bibinfo{author}{Santoro, G.~E.}
\newblock \bibinfo{title}{Floquet resonances close to the adiabatic limit and
  the effect of dissipation}.
\newblock \emph{\bibinfo{journal}{Journal of Statistical Mechanics: Theory and
  Experiment}} \textbf{\bibinfo{volume}{2017}}, \bibinfo{pages}{103104}
  (\bibinfo{year}{2017}).

\bibitem{Note19}
\bibinfo{note}{See Supplemental Material online for T2 characterization
  (Section I) and spin Hamiltonians (Section II) for the three materials,
  calculation of the rotating frame Hamiltonian and torque considerations
  (Section III), quantum protection for different initial states (Section IV)
  and power calibration of the setup (Section V).}

\bibitem{Morton2008}
\bibinfo{author}{Morton, J.}, \bibinfo{author}{Tyryshkin, A.},
  \bibinfo{author}{Brown, R.} \& \bibinfo{author}{Shankar, S.}
\newblock \bibinfo{title}{Solid-state quantum memory using the {{31P}} nuclear
  spin}.
\newblock \emph{\bibinfo{journal}{Nature}} \textbf{\bibinfo{volume}{455}},
  \bibinfo{pages}{1085--1088} (\bibinfo{year}{2008}).

\bibitem{Johansson2013}
\bibinfo{author}{Johansson, J.}, \bibinfo{author}{Nation, P.} \&
  \bibinfo{author}{Nori, F.}
\newblock \bibinfo{title}{{{QuTiP}} 2: {{A Python}} framework for the dynamics
  of open quantum systems}.
\newblock \emph{\bibinfo{journal}{Computer Physics Communications}}
  \textbf{\bibinfo{volume}{184}}, \bibinfo{pages}{1234--1240}
  (\bibinfo{year}{2013}).

\bibitem{Yue2017}
\bibinfo{author}{Yue, G.} \emph{et~al.}
\newblock \bibinfo{title}{Sensitive spin detection using an on-chip
  {{SQUID}}-waveguide resonator}.
\newblock \emph{\bibinfo{journal}{Applied Physics Letters}}
  \textbf{\bibinfo{volume}{111}}, \bibinfo{pages}{202601}
  (\bibinfo{year}{2017}).

\bibitem{Probst2017}
\bibinfo{author}{Probst, S.} \emph{et~al.}
\newblock \bibinfo{title}{Inductive-detection electron-spin resonance
  spectroscopy with 65 spins/ {{Hz}} sensitivity}.
\newblock \emph{\bibinfo{journal}{Applied Physics Letters}}
  \textbf{\bibinfo{volume}{111}}, \bibinfo{pages}{202604}
  (\bibinfo{year}{2017}).

\bibitem{Toida2019}
\bibinfo{author}{Toida, H.} \emph{et~al.}
\newblock \bibinfo{title}{Electron paramagnetic resonance spectroscopy using a
  single artificial atom}.
\newblock \emph{\bibinfo{journal}{Communications Physics}}
  \textbf{\bibinfo{volume}{2}}, \bibinfo{pages}{1--7} (\bibinfo{year}{2019}).

\bibitem{Bertaina2009}
\bibinfo{author}{Bertaina, S.} \emph{et~al.}
\newblock \bibinfo{title}{Multiphoton {{Coherent Manipulation}} in
  {{Large}}-{{Spin Qubits}}}.
\newblock \emph{\bibinfo{journal}{Physical Review Letters}}
  \textbf{\bibinfo{volume}{102}}, \bibinfo{pages}{50501--50504}
  (\bibinfo{year}{2009}).

\bibitem{Smith1959}
\bibinfo{author}{Smith, W.~V.}, \bibinfo{author}{Sorokin, P.~P.},
  \bibinfo{author}{Gelles, I.~L.} \& \bibinfo{author}{Lasher, G.~J.}
\newblock \bibinfo{title}{Electron-{{Spin Resonance}} of {{Nitrogen Donors}} in
  {{Diamond}}}.
\newblock \emph{\bibinfo{journal}{Physical Review}}
  \textbf{\bibinfo{volume}{115}}, \bibinfo{pages}{1546--1552}
  (\bibinfo{year}{1959}).

\bibitem{Wyk1997}
\bibinfo{author}{van Wyk, J.~A.}, \bibinfo{author}{Reynhardt, E.~C.},
  \bibinfo{author}{High, G.~L.} \& \bibinfo{author}{Kiflawi, I.}
\newblock \bibinfo{title}{The dependences of {{ESR}} line widths and spin -
  spin relaxation times of single nitrogen defects on the concentration of
  nitrogen defects in diamond}.
\newblock \emph{\bibinfo{journal}{Journal of Physics D: Applied Physics}}
  \textbf{\bibinfo{volume}{30}}, \bibinfo{pages}{1790--1793}
  (\bibinfo{year}{1997}).

\bibitem{Bertaina2017}
\bibinfo{author}{Bertaina, S.}, \bibinfo{author}{Yue, G.},
  \bibinfo{author}{Dutoit, C.-E.} \& \bibinfo{author}{Chiorescu, I.}
\newblock \bibinfo{title}{Forbidden coherent transfer observed between two
  realizations of quasiharmonic spin systems}.
\newblock \emph{\bibinfo{journal}{Physical Review B}}
  \textbf{\bibinfo{volume}{96}}, \bibinfo{pages}{024428}
  (\bibinfo{year}{2017}).

\bibitem{Bertaina2011}
\bibinfo{author}{Bertaina, S.}, \bibinfo{author}{Groll, N.},
  \bibinfo{author}{Chen, L.} \& \bibinfo{author}{Chiorescu, I.}
\newblock \bibinfo{title}{Multi-photon {{Rabi}} oscillations in high spin
  paramagnetic impurity}.
\newblock \emph{\bibinfo{journal}{Journal of Physics: Conference Series}}
  \textbf{\bibinfo{volume}{324}}, \bibinfo{pages}{012008}
  (\bibinfo{year}{2011}).

\bibitem{Bertaina2015a}
\bibinfo{author}{Bertaina, S.}, \bibinfo{author}{Martens, M.},
  \bibinfo{author}{Egels, M.}, \bibinfo{author}{Barakel, D.} \&
  \bibinfo{author}{Chiorescu, I.}
\newblock \bibinfo{title}{Resonant single-photon and multiphoton coherent
  transitions in a detuned regime}.
\newblock \emph{\bibinfo{journal}{Physical Review B}}
  \textbf{\bibinfo{volume}{92}}, \bibinfo{pages}{024408}
  (\bibinfo{year}{2015}).

\bibitem{Hempstead1960}
\bibinfo{author}{Hempstead, C.} \& \bibinfo{author}{Bowers, K.}
\newblock \bibinfo{title}{Paramagnetic {{Resonance}} of {{Impurities}} in
  {{CaWO4}}. {{I}}. {{Two S}}-{{State Ions}}}.
\newblock \emph{\bibinfo{journal}{Physical Review}}
  \textbf{\bibinfo{volume}{118}}, \bibinfo{pages}{131--134}
  (\bibinfo{year}{1960}).

\end{thebibliography}

\section*{DATA AVAILABILITY}
Data sets generated and analyzed during the current study are available from the
corresponding author on request.

\section*{ACKNOWLEDGEMENTS}
ESR measurements were supported by the CNRS research
infrastructure RENARD (award number IR-RPE CNRS 3443). Partial support by the
National Science Foundation Cooperative Agreement No. DMR-1644779 and the State
of Florida is acknowledged.

\section*{AUTHOR CONTRIBUTIONS}
S.B. and I.C. designed the experiment and analyzed the data. Measurements were performed at Lille University by H.V. and S.B. S.B., I.C. and HDR provided the theoretical background. All authors
contributed to the writing of the manuscript.

\section*{COMPETING INTERESTS}
the Authors declare no Competing Financial or Non-Financial Interests

\pagebreak
\clearpage
\onecolumngrid

\SI

\begin{center}
	\textbf{\large{\textit{Supplementary Information} \\\smallskip
			\bluetitle{Experimental protection of quantum coherence by using a phase-tunable image drive}}}\\
	\author{S. Bertaina}\email{sylvain.bertaina@im2np.fr}\affiliation{\affA}
	\author{H. Vezin}\affiliation{\affC}
	\author{I. Chiorescu} \email{ic@magnet.fsu.edu}\affiliation{\affB}
\end{center}\label{Supplementary}

\section{Carr-Purcell-Meiboom-Gill (CPMG) measurements}
In the absence of inhomogeneous broadening, the dephasing time or transverse
relaxation time $T_2^*$ is directly determined by the ESR linewidth. In solids,
and in particular for single crystals, the anisotropic interactions as well as
their distribution throughout the crystal induce an inhomogeneity of the line.
To measure the dephasing time, the most simple pulse sequence is the Hahn
primary echo: a $\pi/2$ pulse rotates the spins in a plane transverse to the
static field; due to field inhomogeneity, spin will defocus and spread within
the transverse field. A subsequent $\pi$ pulse reverses spin motion and creates
the observed echo signal when refocusing is achieved. By characterizing the
exponential decay of the echo signal as a function of the time between the first
and the second pulse, one obtains the dephasing time $T_2$.

\begin{figure}[h]
	\centering
	\includegraphics[width=0.48\textwidth]{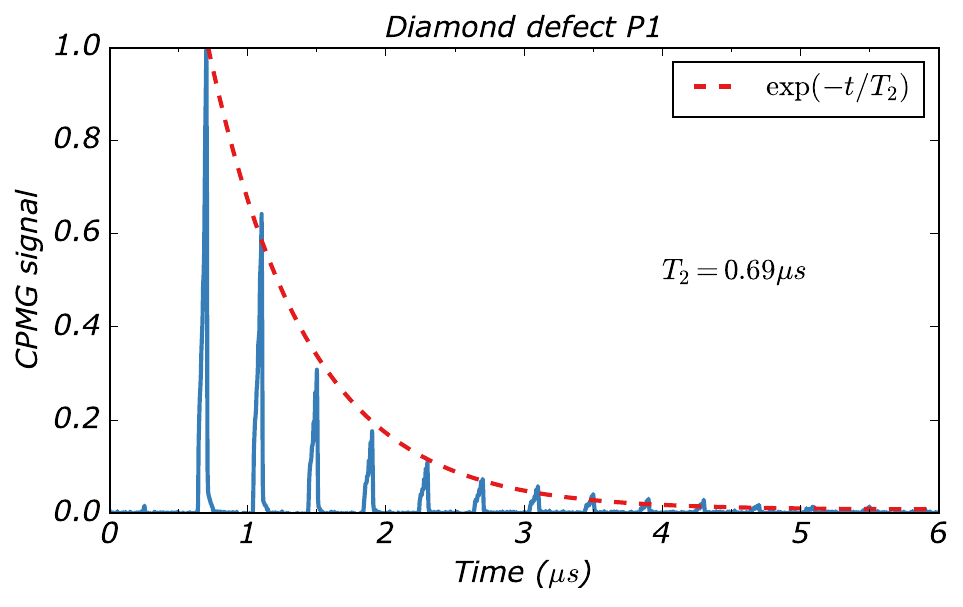}
	\includegraphics[width=0.48\textwidth]{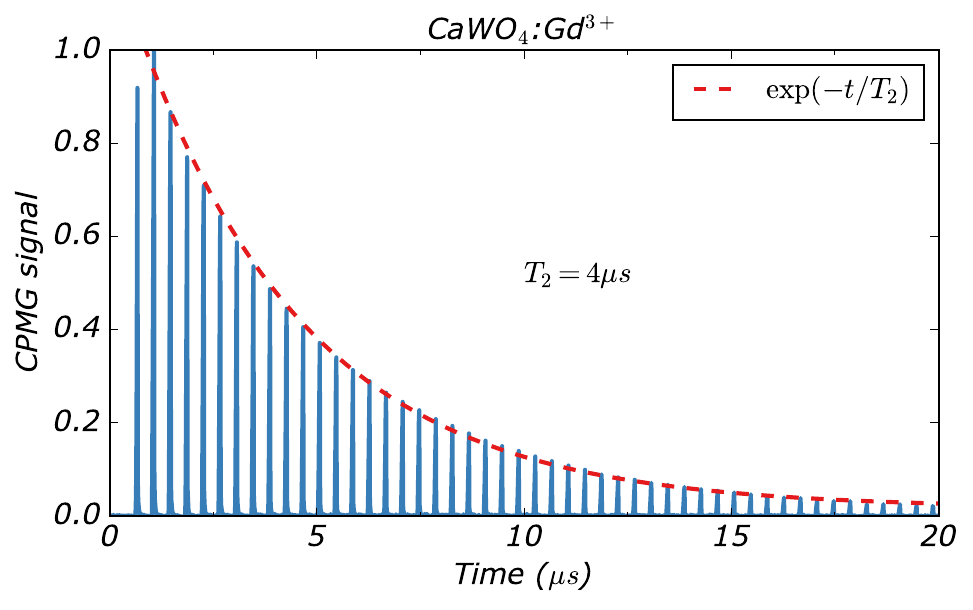}
	\caption{\textbf{Coherence times measured with the CPMG protocol.} 
		CPMG signal normalized to the maximum of the echo intensity (blue line)  as a
		function of the total time of the pulse sequence; each peak denotes the
		addition of a $\pi$-pulse into the sequence. The peaks decay following an
		exponential (red dashed line) with characteristic time $T_2$. (a) Diamond
		defect P1 with $T_2=$~0.69$~\mu$s at T = 15~K. (b) Gd$^{3+}$ spins with
		$T_2=$~4$~\mu$s at T = 40~K.  }
	\label{FigSI:CPMGDiam}
\end{figure}

However, because of diffusion mechanisms, imperfection of pulses or microwave
field inhomogeneity, the refocusing is not complete and the intrinsic dephasing
time measured by this sequence is under-evaluated. To reduce these unwanted
effects, we used the dynamical decoupling offered by the
Carr-Purcell-Meiboom-Gill (CPMG) sequence: after a $(\frac{\pi}{2})_x$ pulse applied along
$x$, a train of $\pi$ pulses is applied along $y$ with alternate
orientations ($\pi_{y},\pi_{-y}$, \ldots). Many primary echoes are thus
generated, with an intensity decreasing exponentially as a function of time and
characterized by the intrinsic dephasing time $T_2$. The measurements of $T_2$
using CPMG sequence for the diamond and \Gd are given in
Fig.~\ref{FigSI:CPMGDiam}, panels (a) and (b) respectively. Measurements are
done at the same temperatures as for the data in the main article Fig.~\1{1} and
lead to a $T_2$ of 0.69~$\mu$s and 4~$\mu$s respectively. A similar result is obtained if the train of $\pi$ pulses has alternate orientations along $x$ and $y$ axes (($\pi_{x},\pi_{y},\pi_{-x},\pi_{-y}$, \ldots).

In the case of \Mn, the inhomogenous absorption linewidth is only 0.05 mT and
the corresponding spin-echo measurement gives a decoherence time of 3~$\mu$s
(Fig.~\1{2}). Since the absorption linewidth is very narrow, the spin-echo
signal of \Mn has a small amplitude and thus subsequent CPMG pulses are not
helpful in this case.

\section{Spin systems characteristics}

The methodology presented here is applied to different
spin systems: the nitrogen substitution in diamond P1 defect ($S=1/2$)
(concentration :100~ppm), Mn$^{2+}$ impurities in MgO ($S=5/2$) with a
concentration of 10~ppm and Gd$^{3+}$ impurities in CaWO$_4$ ($S=7/2$) with a
concentration of 50~ppm.  Despite the large Hilbert space of the Mn$^{2+}$ and
Gd$^{3+}$ spin Hamiltonians, the orientation of the magnetic field and the
frequency and power of the microwave excitation are chosen to avoid multiple
level transitions and thus select only one resonance \cite{Bertaina2009}.
Therefore, the spin systems can be considered as effective two-level systems
undergoing coherent Rabi rotations. The spin Hamiltonians, operating parameters
(fields and frequencies) as well as characteristic $T_{1,2}$ times for these
materials are given below.

\subsection{P1 defects in diamond}
\begin{figure}[h]
	\centering
	\includegraphics[width=0.48\textwidth]{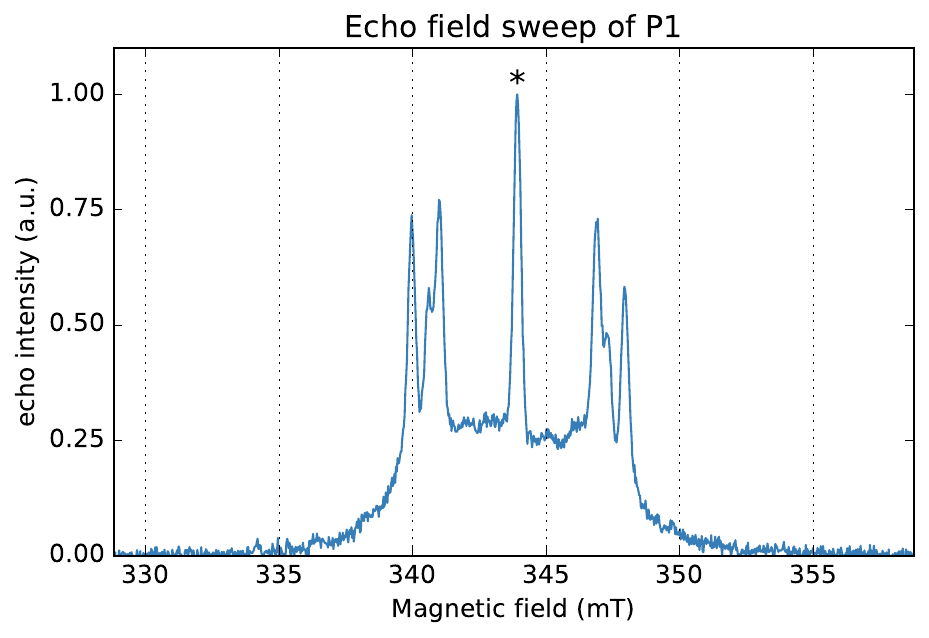}
	\caption{\textbf{Echo field sweep of P1}. Echo signal recorded as a function of static field at T=15~K. The most intense line ($m_I=0$, shown with a star symbol) is not affected by cristal orientation and was selected for the study presented here.  }
	\label{FigSI:P1EFS}
\end{figure}

The substituting nitrogen in diamond has covalent bonds to three surrounding carbons and leaves an unpaired electron on the fourth one, giving rise to a spin $S=1/2$ : this is the P1 defect. Its spin Hamiltonian is given by \cite{Smith1959}:
\begin{equation}
H_{P1}=g_{P1}\mu_B \vec{B}_0\vec{S}+\vec{I}[A_N]\vec{S}
\end{equation}
where $g_{P1}=2.0024$ is the g-factor, $I=1$ is the nuclear spin of $^{14}$N and $[A_N]$ is its hyperfine tensor with $A_\perp=81$~MHz and $A_\|=114$~MHz. A measurement of echo signal as a function of field is shown in Fig. \ref{FigSI:P1EFS}. We studied the central line, corresponding to $m_I=0$, which is not affected by the orientation of the crystal and has the strongest signal. In our experiments, the temperature was set at 15~K, the external field at $B_0=343.62$~mT and the microwave frequency at $f_0=9.645$~GHz. The concentration of nitrogen is about 100~ppm, which leads to values of the linewidth \1{2$\Gamma$=4~G} and dephasing time $T_2=0.69\mu s$, similar to values presented in Ref. \cite{Wyk1997}.

\subsection{\Mn}
\Mn is the second studied system. The non-magnetic matrix of MgO contains a very low concentration ($\sim$ 10 ppm) of $^{55}$Mn$^{2+}$ spins with $S=I=5/2$. The spin Hamiltonian is given by \cite{Bertaina2017}:
\begin{equation}
H_{Mn}=H_{CF}+g_{Mn}\mu_B \vec{B}_0\vec{S}-A\vec{S}\vec{I}
\end{equation}
where $g_{Mn}=2.0014$ is the g-factor, $A=244$~MHz is the hyperfine constant and $H_{CF}$ is a crystal field term resulting from the cubic symmetry $F_{m\bar3 m}$ of MgO: $H_{CF}=a/6[S_x^4 + S_y^4 + S_z^4-S(S + 1)(3S^2 + 1)/5]$. In previous studies \cite{Bertaina2009,Bertaina2011,Bertaina2015a,Bertaina2017} we detailed the effect of $H_{CF}$ on the spin eigenvalues, and showed that the static field orientation can tune in-situ their values to be perfect equidistant or non-harmonic. In the present study we operate the static field under an alignment generating sufficient non-harmonicity such that one can resonantly select two of the six levels and consider the Mn$^{2+}$ a two-level system. Because of the cubic symmetry and the absence of nuclear spin in the host matrix, the line is very narrow with 2$\Gamma=0.5$~G. In this system the detection of $\langle S_z\rangle$ is not done by echo, but by integrating the area under the Free Induction Decay (FID) signal, measured immediately after $\tau_{wait}$ and one $\pi/2$ readout pulse.

The measurements were performed at a temperature of 40~K, with the external field at $B_0=353.7$~mT and the microwave frequency $f_0=9.734$~GHz.

\subsection{\Gd}
\begin{figure}[h]
	\centering
	\includegraphics[width=0.48\textwidth]{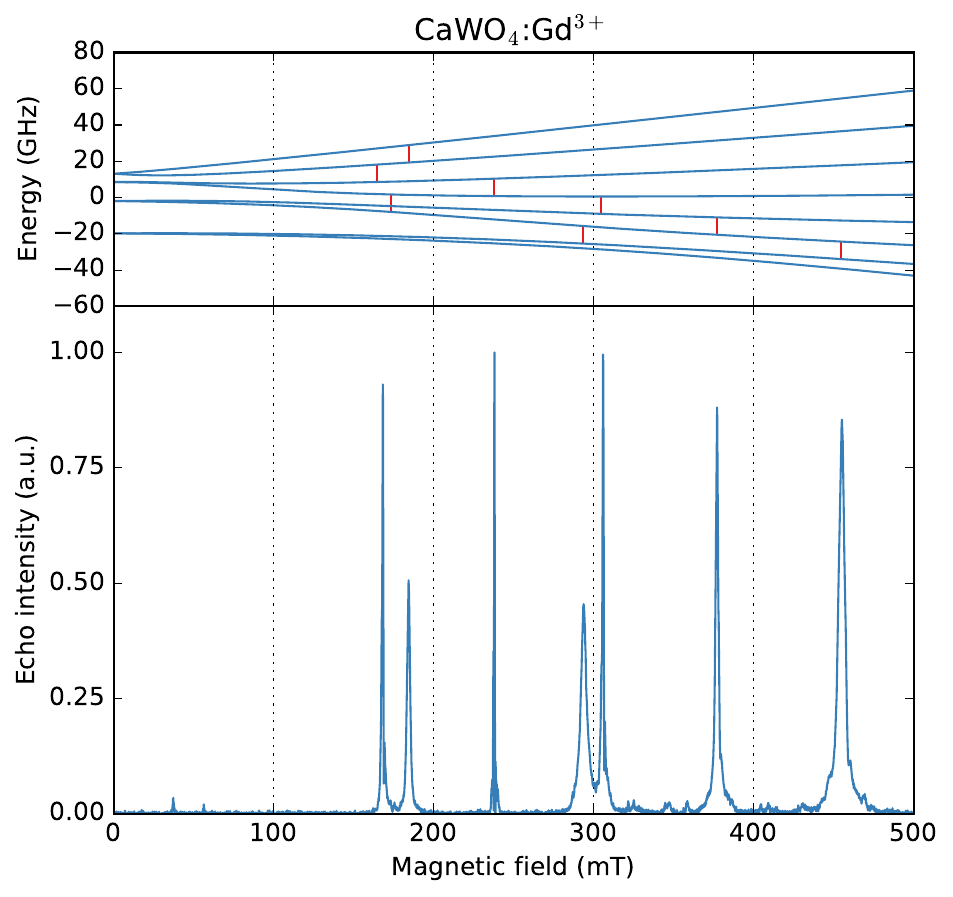}
	\caption{\textbf{Absorption spectra for \Gd.} (top) Energy levels of the Gd$^{3+}$ spin as a function of static magnetic field $B_0$. The red segments indicate level splittings equal to $f_0$. (bottom) Absorption intensity spectra (in arbitrary units) as a function of magnetic field obtained by echo field sweep. Peaks appears at the location of the red segments in the top panel. Their intensity depends on the probability to have a transition between the two levels connected by the red segment. }
	\label{FigSI:GdLevelEFS}
\end{figure}

\Gd is the third studied system. Gd$^{3+}$ spins $S=7/2$ are diluted in non-magnetic matrix of CaWO$_{4}$ and described by the spin Hamiltonian \cite{Hempstead1960,Yue2017} :
\begin{equation}
H_{Gd}=g_{Gd}\mu_B  \vec{B}_0\vec{S}+B^0_2O^0_2+B^0_4O^0_4+B^4_4O^4_4+B^0_6O^0_6+B^4_6O^4_6
\end{equation}
where $g_{Gd}=1.991$ and $B^0_2=-916$, $B^0_4=-1.14$, $B^4_4=-7.02$, $B^0_6=-5.94\times 10^{-4}$ $B^4_6=4.77$ are in MHz units. In our experiments, the temperature was set at 40~K, the external field at $B_0=377.0$~mT along to the crystallographic $a$-axis (tetragonal symmetry $I4/a$) and the microwave frequency at $f_0=9.633$~GHz . The full linewidth is $2\Gamma=$6~G A precise analysis of the resonance fields shows a misalignment of about 2.7$^\circ$. The crystal field anisotropy ensures that the Zeeman levels are not equally spaced, similarly to the case of \Mn discussed above. The eigenvalues of $H_{Gd}$ are presented in Fig.~\ref{FigSI:GdLevelEFS} (top). The bottom panel shows a typical intensity absorption spectrum at constant frequency as a function of $B_0$. Peaks appear at resonance fields, indicated by the red segments in the Zeeman diagram (top) which also show the two levels selected for Rabi oscillations. Thus, it is evident that only two levels are involved in the spin dynamics, making the system an effective TLS. Similar type of spectra are measured for the other two samples (diamond and Mn) in order to select the value of the resonance field. The drive and imaging methodology presented here are independent on which resonance is selected .

\section{Coherent pulses in rotating frame: linear Rabi drive and circularly polarized qubit protection}
\begin{figure}[h]
	\includegraphics[width=3.37 in]{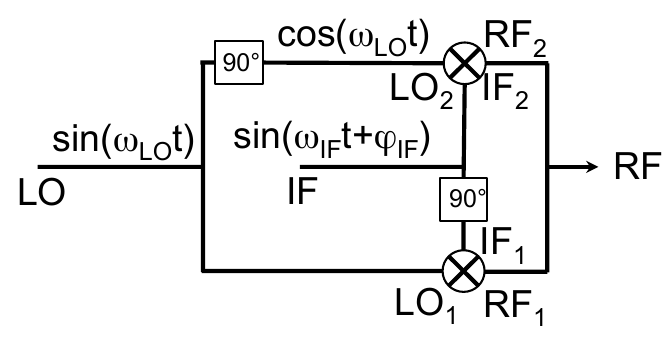}
	\caption{\textbf{Hartley mixer.} A schema of a mixer allowing the creation of the Rabi drive $h_d$ and its coherent image $h_i$, as explained in the text. The main RF drive is at the LO port and it is mixed with a low frequency signal sent into the IF port. The resulting signal exits through the RF port.}
	\label{Fig:Hartley}
\end{figure}

\subsection{Generation of coherent pulses using a Hartley mixer}\label{sec:Hartley} 
Coherent image and Rabi drives can be obtained with a Hartley mixer having slightly unbalanced RF ports, similar to the one used in our setup. An example of such mixer is shown in Fig.~\ref{Fig:Hartley} and we will discuss how a circularly polarized image pulse can be constructed. The input radio-frequency (RF) signal is  $LO=2\sin(\omega_{LO}t)$ and therefore each branch will see $LO_1=\sin(\omega_{LO}t)$ and $LO_2=\cos(\omega_{LO}t)$. Similarly, $IF=2A\sin(\omega_{IF}t+\phi_{IF})$. We introduce amplitude and phase mismatches for the IF ports in the following manner: $IF_1=A_1\cos(\omega_{IF}t+\phi_{IF1})$ and $IF_2=A_2\sin(\omega_{IF}t+\phi_{IF2})$ and introduce the notations $A=(A_1+A_2)/2$, $\delta A=(A_1-A_2)/2$, $\phi=(\phi_{IF1}+\phi_{IF2})/2$, $\delta\phi=(\phi_{IF1}-\phi_{IF2})/2$ and $\omega_\pm=\omega_{LO}\pm\omega_{IF}$. The outputs on the two RF ports are:

\begin{equation}
RF_{1,2}=LO_{1,2}\times IF_{1,2}=\frac{A_{1,2}}{2}\left\{\sin(\omega_+t+\phi_{IF1,2})\pm\sin(\omega_-t-\phi_{IF1,2})\right\}
\end{equation}
and thus
\begin{align}
RF=RF_1+RF_2=\frac{A}{2}\left\{\sin(\omega_+t+\phi_{IF1})+\sin(\omega_-t-\phi_{IF1}) + \sin(\omega_+t+\phi_{IF2})-\sin(\omega_-t-\phi_{IF2})\right\}+\nonumber \\
\frac{\delta A}{2}\left\{\sin(\omega_+t+\phi_{IF1})+\sin(\omega_-t-\phi_{IF1}) - \sin(\omega_+t+\phi_{IF2}) + \sin(\omega_-t-\phi_{IF2})\right\}.
\end{align}
We get
\begin{align}
RF=A\cos\delta\phi\sin(\omega_+t+\phi)+\delta A\sin\delta\phi\cos(\omega_+t+\phi)-
A\sin\delta\phi\cos(\omega_-t-\phi)+ \delta A\cos\delta\phi\sin(\omega_-t-\phi).
\label{RFt}
\end{align}
with the second term being neglible in first order. The first term represents the Rabi drive pulse with an amplitude $A_{d}=A\cos\delta\phi$ while the last two show the image pulse which can be rewritten as:
\begin{align}
\tilde{A}_{i}(t)=\frac{\delta A\cos\delta\phi}{\cos\theta}\{-\sin\theta\cos(\omega_-t-\phi) +\cos\theta\sin(\omega_-t-\phi)\}=A_{i}\sin(\omega_-t-\phi-\theta)
\end{align} 
with $A_{i}=\frac{\delta A\cos\delta\phi}{\cos\theta}$ and $\tan\theta=\frac{A\sin\delta\phi}{\delta A\cos\delta\phi}=\frac{\tan\delta\phi}{\delta A/A}$.

\subsection{Effect of the unperfect Hartley mixer on the spin dynamics}

To study the qubit dynamics, we write its Hamiltonian first in the laboratory frame and then in the rotating frame approximation. A static magnetic field $B_z$ gives a Zeeman splitting in resonance with $\hbar\omega_{LO}$, while the microwave excitation is $\propto RF(t)S_x$ with $RF(t)$ given by Eq.~\ref{RFt}. With notations $\Delta=\frac{\omega_{IF}}{2\pi}$, $f_0=\frac{\omega_{LO}}{2\pi}$ and $\omega_{+,-}=\omega_{LO}\pm\omega_{IF}$, we have (in units of $h$):
\begin{equation}
\mathcal H= f_0S_z+2h_{d}S_x\sin(\omega_+t+\phi)+2h_{i}S_x\sin(\omega_-t-\phi-\theta),
\end{equation} 
where $h_{d,i}$, in units of frequency, represent the intensity of the microwave B-fields resulting from the $A_{d,i}$ voltages expressed above; $f_0, h_{d,i}$ and $\Delta$ are in units of MHz. Consequently, $h_{d}$ is the Rabi frequency $F_R=h_{d}$ resulting from the $A_{d}$ drive pulse alone.

In a frame rotating with $\omega_+$, the Hamiltonian (in units of $h$) is transformed into $H_{RF}=U\mathcal H U^\dagger+\frac{i}{2\pi}\frac{\partial U}{\partial t}U^\dagger$ with  $U=e^{i\omega_+tS_z}$. Thus:

\begin{equation}
\frac{i}{2\pi}\frac{\partial U}{\partial t}U^\dagger=-\frac{\omega_+}{2\pi}S_z
\end{equation}
\begin{equation}
US_xU^\dagger=S_x\cos(\omega_+t)-S_y\sin(\omega_+t).
\end{equation}

This leads to:
\begin{align}
H_{RF}=-\Delta S_z+2h_{d}S_x\sin(\omega_+t+\phi)\cos(\omega_+t)- 2h_{d}S_y\sin(\omega_+t+\phi)\sin(\omega_+t)\nonumber\\
+2h_{i}S_x\sin(\omega_-t-\phi-\theta)\cos(\omega_+t)
-2h_{i}S_y\sin(\omega_-t-\phi-\theta)\sin(\omega_+t)
\end{align}
and, after rejecting the high frequency terms in $2\omega_+$ and $\omega_++\omega_-$,
\begin{equation}
H_{RF}=-\Delta S_z+h_{d}(S_x\sin\phi-S_y\cos\phi)-h_{i}[S_x\sin(4\pi\Delta t+\phi+\theta)+S_y\cos(4\pi\Delta t+\phi+\theta)]
\label{Hrf}
\end{equation}

or as the real part $H_{RF}=\Re(\mathcal H_{RF})=(\mathcal H_{RF}+\mathcal H_{RF}^*)/2$ of Hamiltonian:
\begin{equation}
\mathcal H_{RF}=-\Delta S_z+h_{d}S_+e^{-i(\phi-\pi/2)}+h_{i}S_-e^{-i(4\pi\Delta t+\phi+\pi/2+\theta)}
\end{equation}
or
\begin{equation}
\mathcal H_{RF}=-\Delta S_z+S_+[h_{d}e^{-i(\phi-\pi/2)}+h_{i}e^{i(4\pi\Delta t+\phi+\pi/2+\theta)}].
\label{Hcomplex}
\end{equation}

\subsection{Shirley-Floquet formalism}

One can use the Shirley-Floquet formalism as presented in Eq.~(7) of Ref.\cite{Russomanno2017} to further analyze the eigenvalues of $H_{RF}$. The Floquet modes and eigenenergies can be obtained by diagonalizing the operator $\mathcal K=H_{RF}-i\partial_t$ with the matrix form:

\begin{equation}
\begin{bmatrix}
H_0 & H_1 & H_2 & \dots  & \ldots &\ldots  \\
H_{-1} & H_0-f & H_1 & \dots  &\ldots  &\ldots  \\
H_{-2} & H_{-1} & H_0-2f  & \dots  & \ldots &\ldots  \\
\vdots & \vdots & \vdots & \ddots & \vdots & \vdots  \\
\vdots& \vdots & \vdots & \dots  & H_0-n f & \ldots \\
\vdots & \vdots & \vdots & \dots  & \vdots & \ddots 
\end{bmatrix}
\end{equation}
where $H_n=\frac{1}{\tau}\int_{0}^{\tau}H_{RF}(t)e^{i2\pi n f t}dt$, $\tau=f^{-1}$ and $f=2\Delta$.  

$H_{RF}$'s first two terms are time independent and thus form the $H_0$ component, while the third one gives the terms $n=\pm1$: 

\begin{equation}
H_0=-\Delta S_z+\frac{h_d}{2}[S_+e^{-i(\phi-\pi/2)}+S_-e^{i(\phi-\pi/2)}],
\end{equation}
\begin{equation}
H_{\pm 1}=\frac{h_i}{2}S_\mp e^{\mp i(\phi+\pi/2+\theta)}.
\end{equation}

The spin systems studied here behave as two-level systems and therefore one assumes $S=1/2$ in the matrix above:

\begin{equation}
	H_{SF}=\frac{1}{2}\begin{bmatrix}
		\Delta &h_d e^{-i(\phi-\pi/2)} & 0 & 0&0&0&\ldots\\
		h_d e^{i(\phi-\pi/2)} & -\Delta & h_ie^{-i(\phi+\pi/2+\theta)} & 0 &0&0&\ldots\\
		0 & h_ie^{i(\phi+\pi/2+\theta)} & \Delta-4\Delta & h_d e^{-i(\phi-\pi/2)}&0&0&\ldots\\
		0& 0 & h_d e^{i(\phi-\pi/2)} & -\Delta-4\Delta & h_ie^{-i(\phi+\pi/2+\theta)}&0&\ldots\\
		0&0&0& h_ie^{i(\phi+\pi/2+\theta)}&\Delta-8\Delta & h_d e^{-i(\phi-\pi/2)}&\ldots\\
		0&0& 0 &0&h_d e^{i(\phi-\pi/2)} & -\Delta-8\Delta&\ldots\\
		\vdots & \vdots & \vdots & \vdots  & \vdots &\vdots & \ddots
	\end{bmatrix}
	\label{eq_HSF}
\end{equation}
with $\Delta$ and $h_d$ of comparable magnitudes and much larger than $h_i$. In the limit $h_i\rightarrow0$, we can replace the $2\times2$ diagonal blocks with a diagonal $[F_R/2, -F_R/2, F_R/2-2\Delta, -F_R/2-2\Delta,F_R/2-4\Delta, -F_R/2-4\Delta,\dots ]$ and zero-ed out subdiagonals, with:
\begin{equation}
	F_R=\sqrt{\Delta^2+h_{d}^2}.
\end{equation}
The condition for the image pulse $h_i$ to sustain the coherence between even- and odd-numbered quasi-energies is therefore $-F_R/2=F_R/2-n\Delta$ or
\begin{equation}
	F_R=n\Delta \textrm{ and } \sqrt{\Delta^2+h_{d}^2}=n\Delta \textrm{ with }n=2k , k \in N
	\label{Floquet_res}
\end{equation}

In our study, $n=2$ and the coupling is done between the second and third element of the diagonal, via off-diagonal terms containing both $h_d$ and $h_i$. The splitted eigenvalues of the Shirley-Floquet Hamiltonian are $\phi$-independent. Experimentally, the off-diagonal coupling is larger than the linewidth of the Rabi modes, such that a splitting is observed. 

An analytical estimation of $H_{SF}$ eigenvalues can be done using the second order perturbation theory for the 6x6 block shown in Eq.~\ref{eq_HSF}. The determinant $\det[H_{SF}-\1{\lambda}I]=0$ with $I$ the identity matrix and $\lambda$ representing the eigenvalues, is expanded with Mathematica in powers of $h_i$ and equated to zero up to $h_i^4$. For the levels involved in the $n=2$ resonance, the eigenvalues are given by: 
\begin{equation}
	E_\pm=-\Delta\mp\frac{3}{8}h_i+\frac{7h_i^2}{512\Delta}.
\end{equation}
The splitting can thus be estimated as $E_+-E_-=-\frac{3h_i}{4}$.

\subsection{Torque considerations}

\begin{figure}[h]
	\centering
	\includegraphics[width=0.9\textwidth]{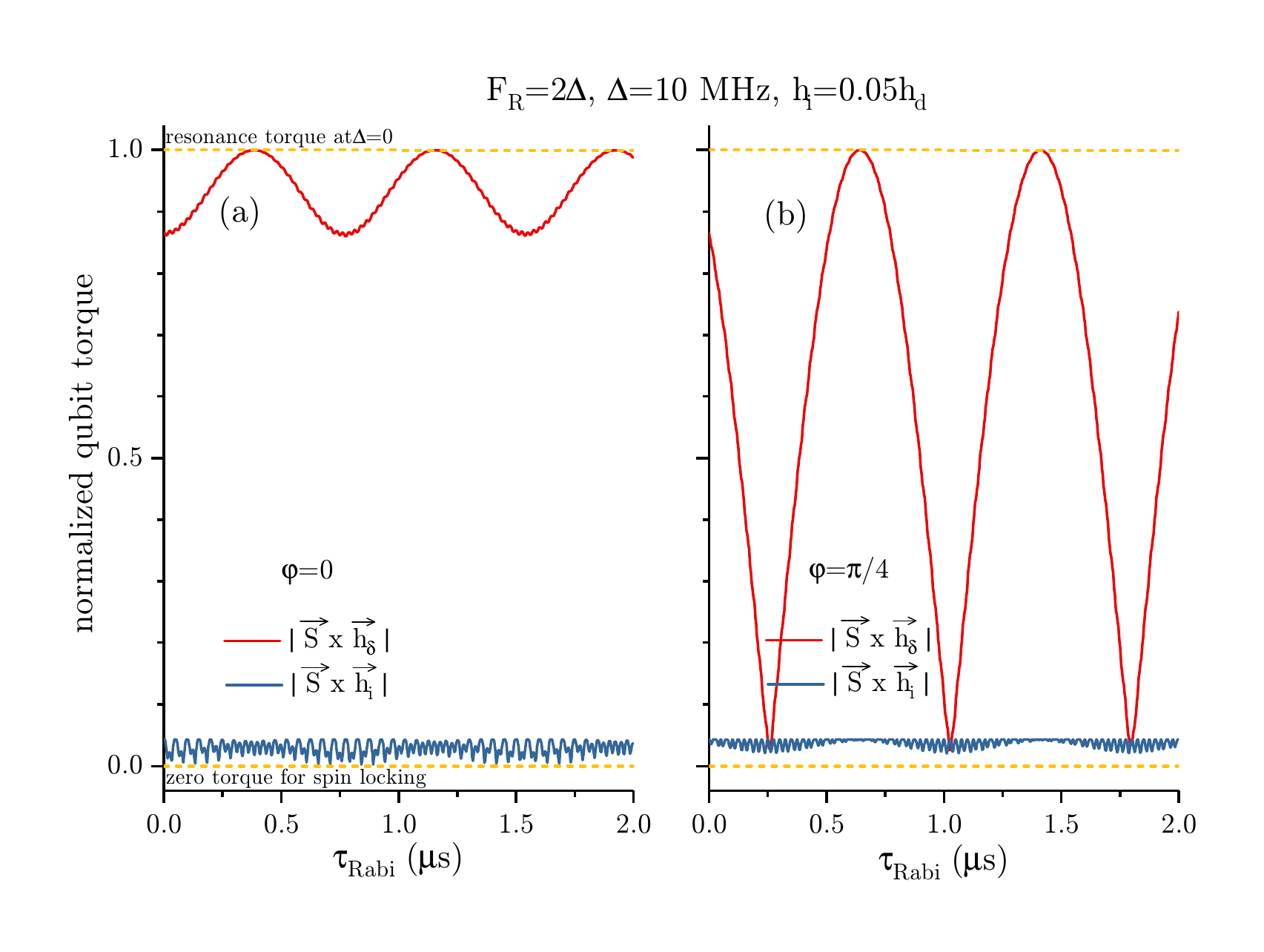}
	\caption{\textbf{Qubit torque during Rabi oscillations.} Torque values are normalized to the maximum value for $\Delta=0$ (top orange dash line) and calculated for the total drive $\vec{h}_\delta$ and the image $\vec{h}_i$ pulses (red and blue lines respectively), starting from the initial ground state. The case of spin-locking is shown by the zero torque dashed line. The qubit dynamics is highly sensitive to the initial phase of the image pulse: (a) $\phi=0$, initial torques are parallel and Rabi oscillations show one main frequency; (b) $\phi=\pi/4$, initial torques are orthogonal and a Floquet mode is generated as a strong beating of the Rabi frequency (see also Fig.~4 of the main article).  }
	\label{FigSI:torque}
\end{figure}

The qubit dynamics imposed by Hamiltonian $H_{RF}$ can be simulated using \texttt{QuTIP}  \cite{Johansson2013} as discussed in the main article. Here, we supplement the understanding of qubit dynamics from the point of view of spin's $\vec{S}$ torque in two situations, $\phi=0$ and $\pi/4$ (see Fig.~\ref{FigSI:torque} \emph{a} and \emph{b}, respectively) in absence of decoherence. Starting from ground state and for a Rabi frequency of 20~MHz, we compute torque magnitudes for the drive and image pulses as a function of time, $|\vec{S}\times\vec{h_\delta}|(t)$ and $|\vec{S}\times\vec{h_i}|(t)$ respectively, with $\vec{h}_\delta=\vec{h}_d+\Delta \hat{z}$ . For simplicity and without loss of generality, the phase $\theta$, an angular shift of the initial phase of $h_i$, is assumed equal to zero.  For comparison, the resonance case at $\Delta=h_i=0$ is shown as an horizontal dashed line with maximum torque value since the spin and the microwave field are orthogonal during the Rabi nutation. All other values are normalized to this maximum value. The zero torque dashed line represents the case of ``spin locking'' when $\vec{S}||h_\delta$. The zero-torque concept of spin-locking applies independently on detuning $\Delta$ but it is valid only if $h_i\equiv0$; otherwise, the image pulse pulls the spin off the axis of $\vec{h}_\delta$. 

In contrast to the spin locking case, both torques are non-zero for our protection protocol. The drive torque at resonance $h_d=\Delta\sqrt{3}$ (in red) is performing the Rabi nutation needed for gate operations while the image torque (in blue) acts as a small perturbation; in this example $h_i$ is 5$\%$ of $h_d$ or -26dB in power. The effect of the initial phase $\phi$ of $h_i$ is essential to qubit dynamics although $h_i$ remains a small perturbation during the gate operation. When the initial torques are parallel or antiparallel ($\phi=0$ or $\pi/2$ respectively) the Rabi rotation is closer to full amplitude and the Floquet mode is less visible. This is shown by a small modulation in drive torque (in red, panel \emph{(a)}) with the period of the Floquet mode. In contrast, when torques are initially perpendicular to each other ($\phi=\pi/4$, panel \emph{(b)}), the Rabi nutation and drive torque has strong beatings, while retaining the same amount of coherence protection. The Floquet mode is clearly visible in this case, as discussed in Fig.~4 of the main article.

\section{Spectrometer setup and field calibration}

\subsection{Spectrometer setup} The measurements have been performed on a conventional
pulse ESR spectrometer Bruker E680 equipped with an incoherent electron double
resonance (ELDOR) bridge and a coherent arbitrary waveform generator (AWG)
bridge. In the ELDOR bridge (Fig. 1\TB{a}), the drive and the
read out pulses come from two independent sources while with the AWG bridge
(Fig.1\TB{b}) all the pulses are generated using the same
microwave source and thus they are all phase coherent.  The drive frequency is
generated by mixing the source $f_0$ (used as a local oscillator) with a low
frequency and phase controllable signal $\DP$ through an in-phase quadrature
(IQ) mixer. Ideally, the output of the mixer is monochromatic with the frequency
$f_0+\Delta$. In reality, the output consists of a principal frequency
$f_0+\Delta$ (the drive) and of lower amplitude images $f_0+n\Delta$. Since the effect of the
image drive is the central part of this paper, we have characterized the AWG bridge
using a spectrum analyzer, right before the power amplification stage. An
example of spectrum is presented in Fig.~\ref{FigSI:awg}. The power of the image drive $f_0-\Delta$ is lower by $\approx -18$~dB
than $f_0+\Delta$. Consequently, an amplitude ratio of the MW magnetic fields
$h_i/h_d$ around $\sim$0.12 is used in simulations.

\subsection{Calibration of the microwave fields} The method presented in this article is highly dependent on the value of the microwave fields $h_d$ and $h_i$. The drive intensity $h_d$ is easy to calibrate by measuring the Rabi frequency at resonance while knowing its expected value from the spin Hamiltonian in the rotating frame ($\Delta=0$).

\begin{figure}[h]
	\centering
	\includegraphics[width=0.48\textwidth]{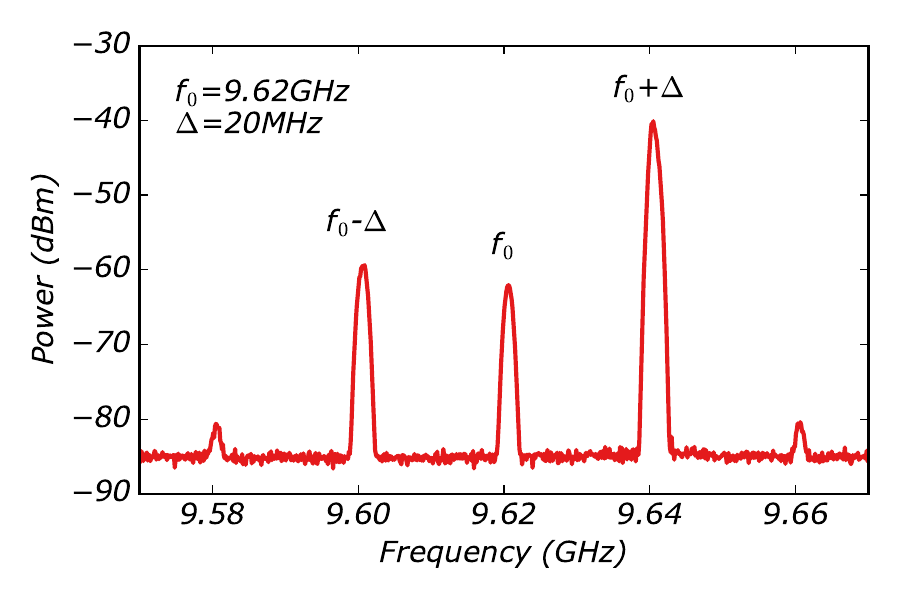}
	\caption{\textbf{Power spectra of the RF signal.} Spectrum analyzer data showing the RF power spectra, after the mixer shown in Fig.~1 of the main article. The Rabi drive pulse is located at $f_0+\Delta$, $\sim18$~dB stronger than the image pulse used for qubit protection, located at $f_0-\Delta$.}
	\label{FigSI:awg}
\end{figure}

On the contrary, $h_i$ is more difficult to calibrate. As seen in Section~\ref{sec:Hartley}, $h_i$ comes from the inherent unbalance of a real mixer and therefore it can be device depended. To calibrate $h_i$, we have used a spectrum analyzer Agilent Technologie, PXA Signal Analyzer N9030A. We have measured the Fourier transform of the microwave coming from the AWG bridge before the power amplification stage as shown in Fig.~\ref{FigSI:awg}: $f_0$ is the carrier frequency, $f_0+\Delta$ is the Rabi drive frequency ($h_d$) and $f_0-\Delta$ is the image frequency. The image pulse is about 100 times ($\sim18$~dB) weaker than the Rabi drive and thus it can be used to sustain the motion rather than driving it.  The signal at $f_0$, about 4-6 dB smaller than the image drive, is not in a Floquet resonance with the Rabi drive (see Eq.~\ref{Floquet_res}) and thus not contributing to spin dynamics.

\end{document}